\documentclass[a4paper,11pt]{article}
\usepackage{jheppub}

\usepackage[T1]{fontenc} 

\usepackage{setspace}
\usepackage{subcaption}
\usepackage{physics}
\usepackage{dsfont}
\usepackage{tensor}
\usepackage[normalem]{ulem}
\usepackage{xcolor}
\usepackage{graphicx}
\usepackage[export]{adjustbox}
\usepackage{mathtools}
\usepackage{mathbbol}
\usepackage{boondox-cal}
\usepackage[centertableaux]{ytableau}
\usepackage{sidenotes}
\usepackage{mathdots}
\usepackage{tikz}

\colorlet{darkgreen}{green!50!black}

\title{APD-Invariant Tensor Networks from Matrix Quantum Mechanics}
   \author[a]{Alexander Frenkel}

   \affiliation[a] {Stanford Institute for Theoretical Physics, Stanford University,\\382 Via Pueblo, Stanford CA 94305}

   \emailAdd{afrenkel@stanford.edu}

   \abstract{
        We propose a simple connection between matrix quantum mechanics and tensor networks. This allows us to imbue tensor networks with some interesting additional structure. The geometry of the graph describing the tensor network state is determined dynamically, giving a notion of background independence. The tensor network states have a $U(N)$ invariance, which (a) allows us to consider continuous families of entanglement cuts even with a finite number of tensors and (b) includes a notion of bulk coordinate reparameterization and area-preserving diffeomorphism invariance in the large $N$ limit. These tensor networks also have a natural scale of nonlocality that behaves similarly to a string scale, suggesting a potential toy model for sub-AdS physics. Emergent $U(p)$ gauge fields naturally appear on the tensor network links.
   }
\date{\today}

\begin{document}

\maketitle

\section{Introduction}\label{sec:intro}

Following the second superstring revolution \cite{schwarz1987review,polchinski1996tasi} a family of holographic dualities was proposed between large $N$ quantum theories and various corners of string theory \cite{banks1999m,Maldacena:1997re,itzhaki1998supergravity,berenstein2002strings,gubser1998gauge,witten1998anti}. These dualities share a heart -- their quantum mechanical sides are generically worldvolume theories of some stack of $N$ D$p$-branes in the decoupling limit. The most well-studied example is the gravitational theory emergent from dimensionally reduced 9+1d $\mathcal{N}=4$ Super Yang-Mills (SYM) \cite{Maldacena:1997re,itzhaki1998supergravity}. The worldvolume theory of D3-branes in particular is conformal, and the emergent spacetime is string theory in AdS$_5 \cross S^5$ \cite{Maldacena:1997re}, so the case $p=3$ is a particular instantiation of the more general AdS/CFT correspondence. However, following the general arguments of \cite{t1993planar}, even non-conformal large-$N$ theories in the appropriate limit often define some emergent string theory. 

\begin{figure}[ht]
     \centering
     \begin{subfigure}[b]{0.45\textwidth}
         \centering
         \includegraphics[width=\textwidth]{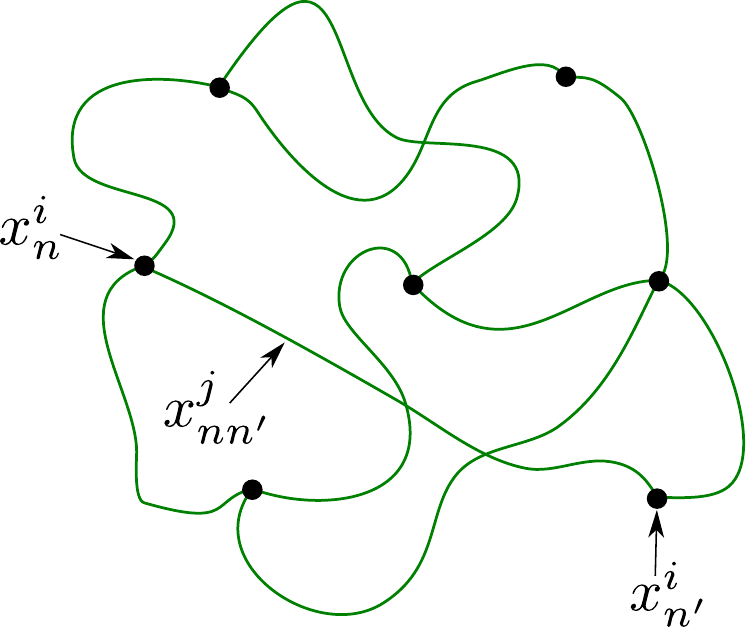}
         \caption{}
         \label{fig:brane-net}
     \end{subfigure}
     \hfill
     \begin{subfigure}[b]{0.45\textwidth}
         \centering
         \includegraphics[width=\textwidth]{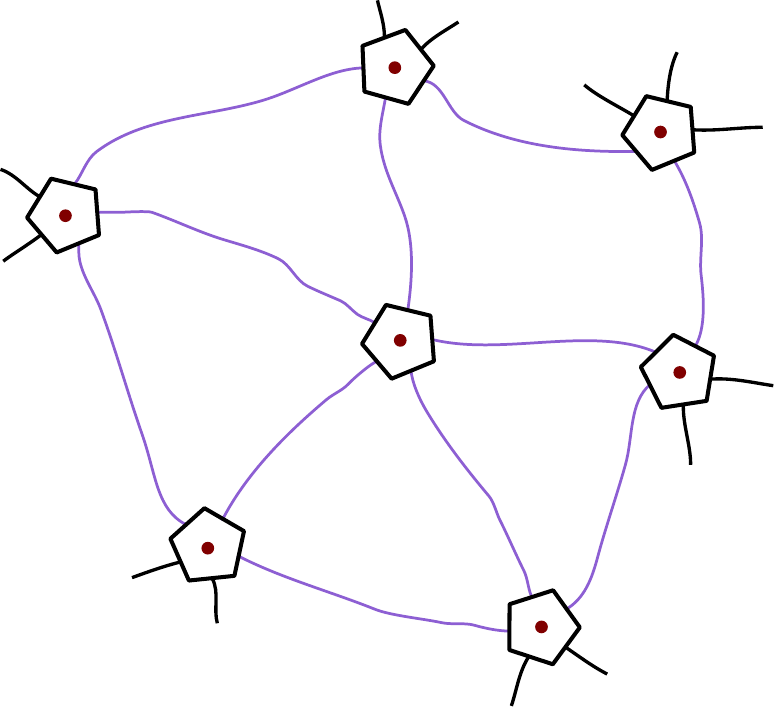}
         \caption{}
         \label{fig:tensor-net}
     \end{subfigure}
     \caption{On the left, in Fig. \ref{fig:brane-net}, we have depicted a collection of D0-branes with their interactions mediated by strings. The positions of the D0-branes are associated to eigenvalues of Hermitian matrices $X^i$, denoted $x^i_n$. The off-diagonal entries of $X^i$, denoted $x^i_{nn'}$, capture the state of string modes stretched between the $n$th and $n'$th branes. On the right, in Fig. \ref{fig:tensor-net}, we have drawn a tensor network state with pentagonal tensors (as in e.g. \cite{pastawski2015holographic}). Contracted legs are drawn in purple, the free external legs associated to a holographic boundary are black, and bulk external legs are depicted by the maroon dots on the tensors. The two cartoons bear a resemblance, and the purpose of this paper is to begin building the technical relationship between them. Similar figures have been drawn in \cite{Hampapura:2020hfg}.}
\end{figure}

In this paper we focus on matrix quantum mechanics (MQM), in which the `boundary' theory is neither conformal, nor a QFT, but a quantum mechanical system with a finite number of degrees of freedom living on a 0+1 dimensional base space. The BFSS \cite{banks1999m} and BMN \cite{berenstein2002strings} models are particular examples of type IIA string theory emergent from interacting D0-branes. The degrees of freedom consist of $N \times N$ Hermitian matrices $X^i$ and fermionic matrices $\lambda^{\alpha}$. The entries of these matrices are denoted $x^i_{nn'}$ and $\lambda_{nn'}^{\alpha}$, with the color indices $n$ running from 1 to $N$. When considered as the worldvolume theory of $N$ D0-branes, the eigenvalues of the matrix $X^i$ are interpreted as the positions of the branes along the $i$th spatial dimension and the off-diagonal elements $x^i_{nn'}$ as modes of strings stretching between the $n$th and $n'$th branes \cite{polchinski1996tasi}. See Fig. \ref{fig:brane-net} for a diagram. As before, MQM Lagrangians with stringy duals have the schematic form of dimensionally reduced SYM:
\begin{equation}\label{eqn:mqm-lag}
L = \Tr[\sum_i \dot{X}^{i 2} + \sum_{ij}[X^i,X^j]^2 + \text{susy terms} + \ldots].
\end{equation}
The commutator squared term may be recognized as the zero-mode piece of the covariant SYM derivative. It is precisely this term that is responsible for imbuing strings with a potential energy linear in their length (see \cite{taylor2001m} for a review).

Since the discovery of holographic duality, there has been a concerted effort to develop tractable toy models that capture important aspects of its physics. Tensor networks \cite{pastawski2015holographic,Hayden:2016cfa,swingle2012entanglement} have been one of the most celebrated and fruitful of these models, and particularly excel in capturing the emergence of bulk geometry from quantum correlation in the boundary theory. Among other phenomena they capture holographic error correction, bulk reconstruction, the Ryu-Takayanagi formula for boundary entropy \cite{ryu2006holographic}, and a notion of sub-AdS locality \cite{yang2016bidirectional}. Suprisingly, they have even recently been shown to recover bulk gravitational forces that appear to directly arise from the entanglement pattern \cite{Sahay:2024vfw}. They are essentially lattice-regularized holography. 

The key building block of of a tensor network is a generic state on $k$ qudits with Hilbert space dimension $d$. The most general such state is described in terms of a multilinear operator (i.e. tensor) $t^{\mu_1\ldots \mu_k}$:
\begin{equation}
\ket{t} = \sum_{\{\mu_i\} = 0}^{d-1} t^{\mu_1\ldots\mu_k}\ket{\mu_1} \otimes \ldots \otimes \ket{\mu_k}.
\end{equation}
These states are pictorially represented as a geometric shape with $k$ external legs, with one qudit associated to each leg (see Fig. \ref{fig:tensor-net} for an example). General tensor network states are built by entangling qudits on the external legs of different tensors. This is done by taking inner products of the tensors $t^{\mu_1 \ldots \mu_k}$. To represent entangled qudits, we connect the legs of the tensors to form an internal edge. As such, tensor network states are graphs (again, see Fig. \ref{fig:tensor-net}).

While they successfully capture many aspects of AdS/CFT, tensor networks suffer from deficiencies. As they are defined on a lattice determined by some graph, they lose the diffeomorphism invariance and background independence that constitute general covariance of the bulk geometry (see \cite{Akers:2024wab} for an alternative recent approach for introducing diffeomorphism invariance using the Chern-Simons description of 3d gravity). It is further unclear how tensor networks properly describe the backreaction of RT surfaces on the geometry \cite{Dong:2023kyr,Akers:2024wab}. While capable of probing some sub-AdS lengthscale physics, they do not capture $\alpha'$ corrections and the UV/IR mixing we expect from physics at the string scale.

In this paper we argue that MQM states are naturally thought of as a generalization of tensor network states. The construction is built on a simple observation -- the Hilbert space of MQM is identical in structure to the Hilbert space of tensor networks. Moreover, the matrices $X^i$ defining compact noncommutative manifolds have a natural interpretation as graph adjacency matrices. Mapping tensor network states into $U(N)$ invariant MQM models allows us to define tensor networks on noncommutative manifolds. The $U(N)$ symmetry in the large $N$ limit contains area-preserving diffeomorphisms (APDs) \cite{hoppe1982quantum} (see \cite{swain2004topology} for discussion of some subtleties on this point), and a judicious choice of Hamiltonian allows the background geometry to be determined dynamically, so we may think of this construction as a step toward introducing general covariance and background independence into tensor network models. This construction may be viewed as a particular instantiation of the ideas in \cite{Cheng:2022ori}.

\subsection{Overview of the Construction}

The construction in this paper relies on results and intuition from tensor networks, noncommutative geometry, and noncommutative field theory. For the convenience of a reader that may unfamiliar with one or more of these concepts we give a brief sketch of the core idea, where we give the essential results but leave their technical justification for \S\ref{sec:matrix-tn}. 

\begin{figure}[ht]
\centering
\includegraphics[width=\textwidth]{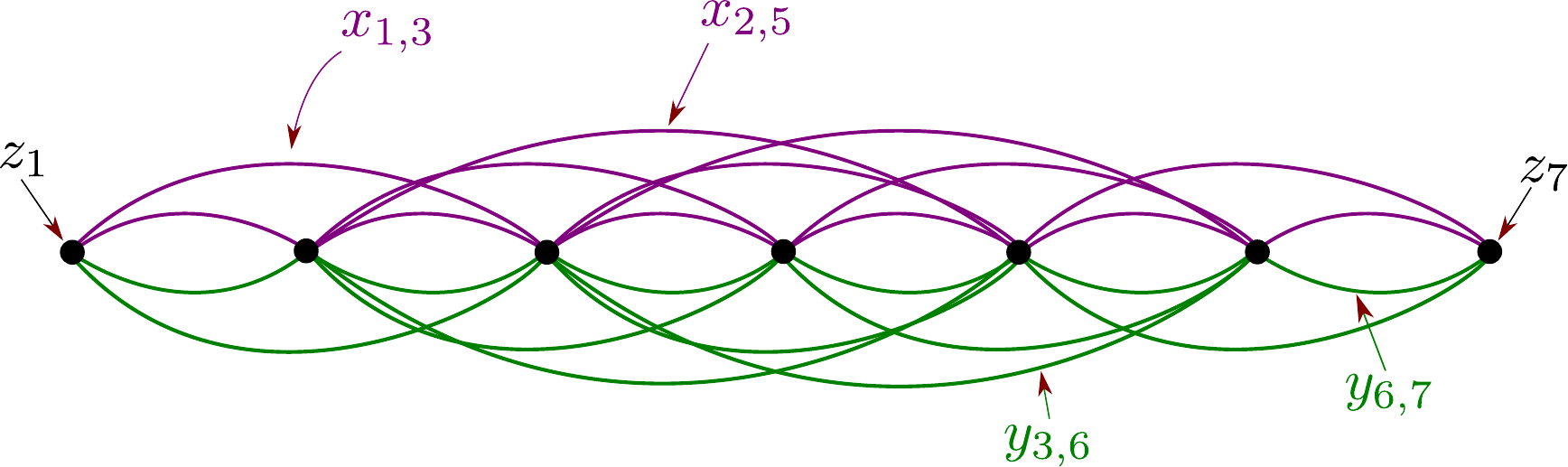}
\caption{A sketch of the graph interpretation of a noncommutative geometry defined by three coordinate matrices $X$, $Y$, and $Z$, in the case $N = 7$. We have chosen to diagonalize $Z$. Each node represents the location of one of the eigenvalues $z_n$ of $Z$. To each nonzero entry of $X$ and $Y$ we assign an edge. $X$ edges are drawn above the nodes and in purple, $Y$ edges are drawn below the nodes and in green. In this example entry $x_{1,3}$ is nonzero and the entry $x_{1,7}$ is zero.}\label{fig:MTN-sketch}
\end{figure}

As we review in \S\ref{sec:fuzzy-review}, a noncommutative manifold $\mathcal{M}_{\theta}$ is defined via a set of $N \times N$ matrices $X^i$. The $X^i$ are the generators of the algebra of functions on $\mathcal{M}_{\theta}$, and are themselves interpreted as the coordinate functions. One of the observations of \S \ref{sec:fuzzy-review} is that $X^i$ carry a natural interpretation as the adjacency matrices of a graph. For similar observations, see \cite{Hampapura:2020hfg,Frenkel:2021yql}. An example of such a graph in the case of $N=7$ is drawn in Fig. \ref{fig:MTN-sketch}.

We wish to put a tensor network on this graph in a manner that respects the $U(N)$ coordinate reparameterization symmetry of noncommutative manifolds. To do so, we introduce $N \times N$ matrices of qubits $\lambda^{\alpha}$. $\alpha$ is a flavor index that runs from $1$ to $N_F$. Our total Hilbert space therefore consists of the $X^i$ and $\lambda^{\alpha}$ degrees of freedom, and there is a natural adjoint action of the symmetry group $U(N)$ on this Hilbert space. The explicit generators of this action are given in \eqref{eqn:U-gens}. In view of Fig. \ref{fig:tensor-net}, maximally entangling the $\lambda_{nn'}^{\alpha}$ qubits with the $\lambda_{n'n}^{\alpha}$ qubits creates a link between the $n$th and $n'$th nodes.
\begin{figure}[ht]
\centering
\includegraphics[width=\textwidth]{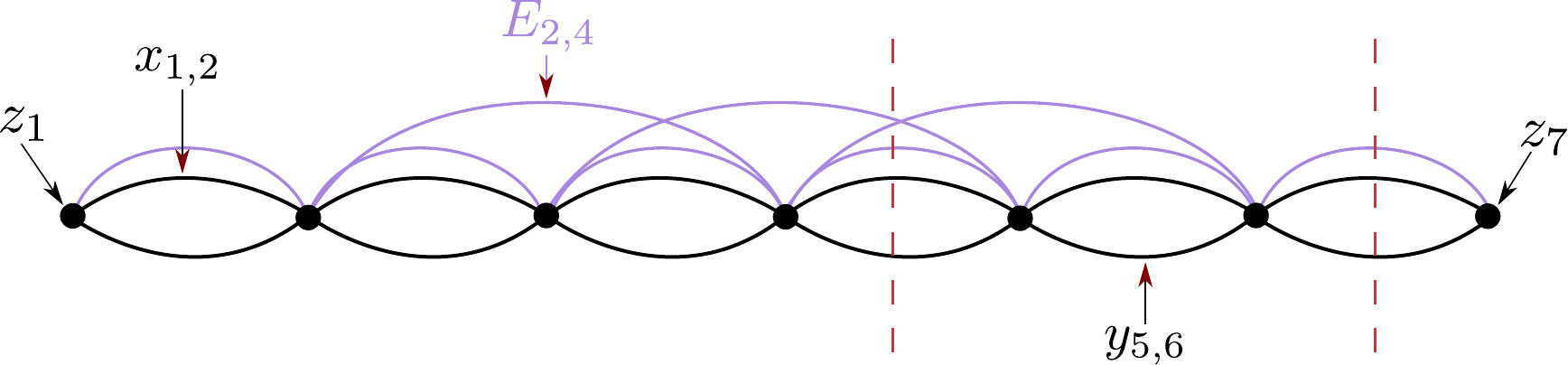}
\caption{A sketch of the tensor network defined on a noncommutative geometry. This graph should be interpreted the same way as Fig. \ref{fig:tensor-net}. The black nodes and edges represent the geometry as in Fig. \ref{fig:MTN-sketch}. On this geometry only nearest neighbor nodes are connected, such as in the fuzzy sphere \eqref{eqn:fuzz-sphere}. The colored edges represent entanglement between qubits on the corresponding nodes -- for example, the edge $E_{2,4}$ represents a maximally entangled state between $\lambda^{\alpha}_{2,4}$ and $\lambda^{\alpha}_{4,2}$. These colored edges are the same objects as the edges of the graph and Fig. \ref{fig:tensor-net}. The vertical dashed lines are two different choices of entanglement cut. The entanglement across these cuts is determined by the number of colored edges they intersect. The entanglement across the leftmost cut is $3N_f\log 2$, and the entanglement across the rightmost cut is $N_f \log 2$. For a more geometric depiction of this network see Fig. \ref{fig:network-apds-intro}. It may seem that we may only partition this network along the $z$ coordinate, but it turns out the action of $U(N)$ transformations on the $X^i$ and $\lambda^{\alpha}$ matrices correspond to area-preserving diffeomorphisms that allow us to consider any possible subregion of the geometry (again see Fig. \ref{fig:network-apds-intro}).}\label{fig:MTN-sketch-2}
\end{figure}

The key to the construction is how we determine the state of the tensor network. We write $U(N)$-invariant MQM Hamiltonians on the $X^i$ and $\lambda^{\alpha}$ degrees of freedom and consider their low energy states. While we expect the relationship between MQM and tensor networks to be quite general, for simplicity the specific Hamiltonians we consider in this paper take the form
\begin{align}
&H = H_{geom.} + \kappa \left(H_1 + H_2\right),\label{eqn:full-ham-intro}\\
&H_{geom.} = \Tr[\sum_i {\Pi^i}^2+ V(X^i)],\\
&H_1 = i\sum_{\alpha}\Tr[(\lambda^{\dag \alpha})^2 - (\lambda^{\alpha})^2]\\
&H_2 = \sum_{\alpha} \Tr[\lambda^{\alpha \dag}(g_1[X^i,[X^i,\lambda^{\alpha}]] + g_2[X^i,[X^i,[X^j,[X^j,\lambda^{\alpha}]]]] + \ldots)].
\end{align}
$H_{geom.}$ is the Hamiltonian that sets the geometry of the noncommutative manifold. It consists of a kinetic term (the $\Pi^i$ are the conjugate momenta of the $X^i$) and a potential term $V(X^i)$. This could be something like mini-BMN \cite{Anous:2017mwr,Han:2019wue}, which has the fuzzy sphere as a ground state. $H_1$ is designed to maximally entangle all $\lambda_{nn'}^{\alpha}$ qubits with their $\lambda_{n'n}^{\alpha}$ counterparts. $\kappa$ controls the strength of the backreaction of the fermion modes $\lambda^{\alpha}$ on the geometry. For the purposes of this paper we work in the regime of $\kappa \ll 1$, so as not to worry about this backreaction. 

The ground state of $H_1$ is an all-to-all connected tensor network graph. $H_2$ is the coupling between the geometry and the tensor network qubits. On its own, the ground state of $H_2$ is the completely unentangled state of all the $\lambda^{\alpha}$ qubits -- without $H_1$, this would define a completely disconnected network. The commutator structure has a nice interpretation in the noncommutative geometry -- the commutator $[X^i,[X^i,\cdot]]$ is the noncommutative analog of the Laplacian $\Delta$. We should therefore read $H_2$ as
\begin{equation}
H_2 \sim \int_{\mathcal{M}_{\theta}} \sqrt{g}d^D\sigma\, \lambda^{\dag}\left(g_1\Delta \lambda + g_2 \Delta^2 \lambda + \ldots\right),
\end{equation}
where the couplings $g_i$ set a derivative expansion in the Hamiltonian of some spinor fields $\lambda^{\alpha}(\sigma)$ propagating on $\mathcal{M}_{\theta}$. The structure of $H_2$ suppresses high momentum modes of $\lambda^{\alpha}$, and by tuning $g_i$ we may set a momentum scale for modes that can propagate in the low-energy subspace of \eqref{eqn:full-ham-intro}. The nature of noncommutative geometry is that momentum scales are related to length-scales -- high momentum modes spread out in the transverse directions to the direction of propagation \cite{McGreevy:2000cw}. By setting a momentum scale, we also set a length-scale across which internal edges of the tensor network can stretch. This is the scale of nonlocality of the network. The competition between $H_1$ and $H_2$ is what sets the scale of nonlocality and creates the local entanglement strucure of the states we consider.

\begin{figure}[ht]
\centering
\includegraphics[width=\textwidth]{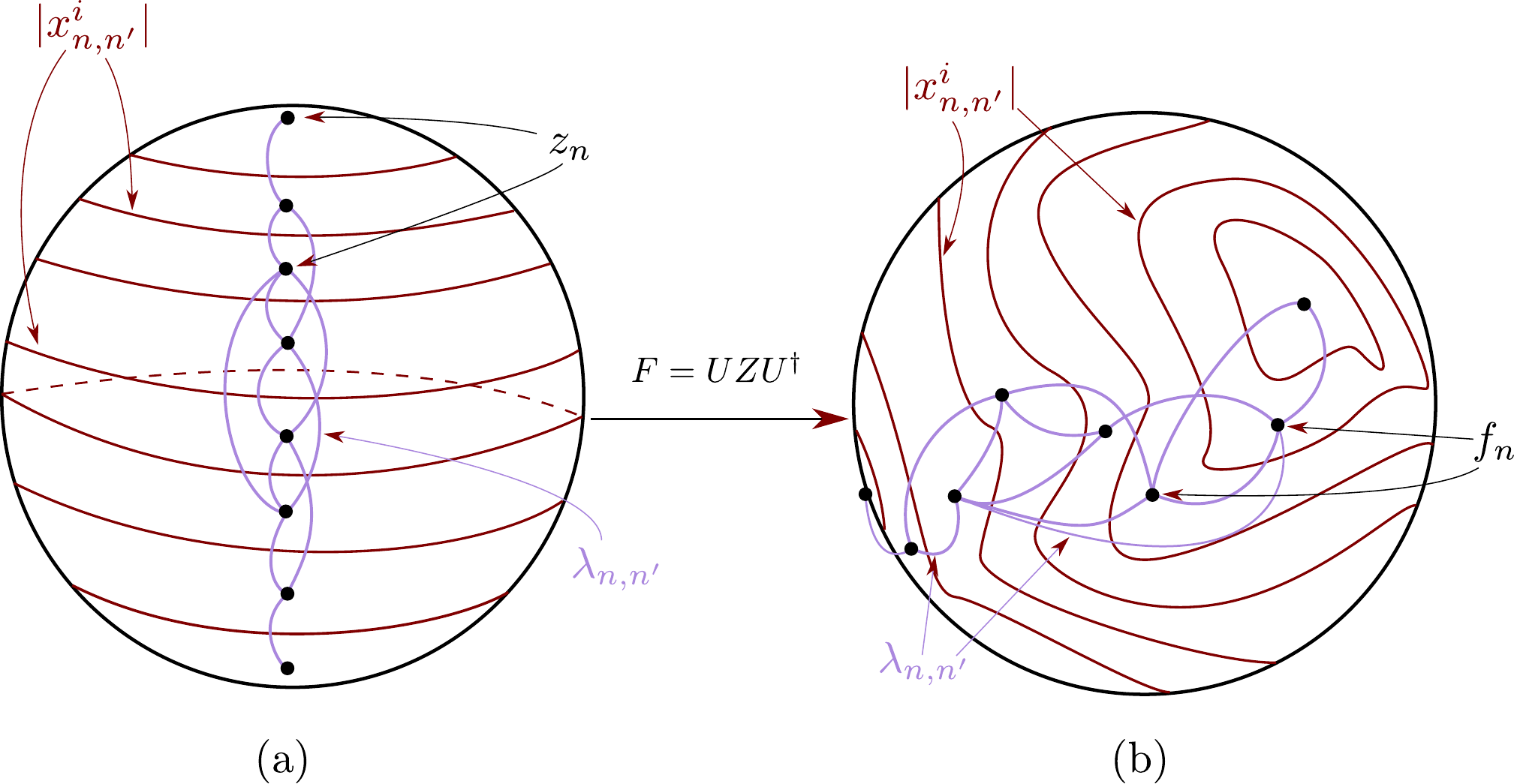}
\caption{In this figure we depict how the tensor network drawn in Fig. \ref{fig:MTN-sketch-2} lives on a noncomutative geometry. We have chosen the fuzzy sphere for simplicity. In (a) we have chosen to diagonalize the $Z$ coordinate, so each node of the network is associated to a region of area $1/N$ localized to some $z = z_n$. The entries of the other matrix entries $X^i$ that are not diagonal are closely related by the lengths of the borders between these regions in a manner described in \S\ref{sec:fuzzy-review}. To go from (a) to (b), we have performed a change of basis implemented by some unitary $U \in U(N)$. In the new basis a different curvilinear coordinate $F = f(X,Y,Z)$ is diagonalized, so now each area $1/N$ region is localized to some $f(x,y,z) = f_n$. Again, the entries of $X^i$ are related to the lengths of the borders between these regions. This is the noncommutative geometry analog of an area-preserving diffeomorphism (APD) (see the discussions in \cite{hoppe1982quantum,Frenkel:2021yql,swain2004topology}). The same unitary $U$ that implements this APD also transforms the state of the fermions $\lambda^{\alpha}$, so the tensor network state transforms covariantly with this coordinate reparameterization. In both (a) and (b) the total number of tensor network edges (i.e. sets of entangled $\lambda^{\alpha}_{nn'}$ qubits) crossing each boundary between regions is equal to the length of this boundary, so the area law entanglement properties of the tensor network are preserved in any choice of coordinate system.}\label{fig:network-apds-intro}
\end{figure}

It is crucial that the Hamiltonian of \eqref{eqn:full-ham-intro} is $U(N)$ invariant, so that we may restrict ourselves to the subspace of $U(N)$-invariant states (i.e. we may gauge this symmetry\footnote{In quantum mechanics, as opposed to quantum field theory, we often have a choice about whether or not to take a symmetry to be global symmetry or a gauge redundancy. It is only a question of whether we restrict to the subspace of the Hilbert space annihilated by the symmetry generators.}). So far, as depicted in Fig. \ref{fig:MTN-sketch-2}, we have described tensor networks where each node has a fixed $z$ coordinate. It may seem that this only allows us to partition our network along $z$. However, by acting with a unitary transformation on the $X^i$ and $\lambda^{\alpha}$ matrices, we may transform the state of our tensor network to a new basis that admits a Hilbert space factorization along any curvilinear coordinate\footnote{There is a UV cutoff on the curvature of coordinates we may consider given by $1/N$, the scale of noncommutativity. However, it is important to note that unlike a standard tensor network, we may consider continuous families of coordinates (and therefore a continuous families of entanglement cuts).} $f(x)$ of our choosing (see Fig. \ref{fig:network-apds-intro}).

A punchline here is that in the large $N$ limit, area-preserving diffeomorphisms are a gauge symmetry of these matrix tensor networks. Different $U(N)$ gauge fixing conditions behave as different coordinate systems on $\mathcal{M}_{\theta}$, and each coordinate system presents us with a different coordinate along which we may partition the network and compute an entanglement entropy (again see Fig. \ref{fig:network-apds-intro}). It is quite nontrivial that the ground state of \eqref{eqn:full-ham-intro} has area-law entanglement for any choice of entanglement cut\footnote{The cut should be weakly curved compared to the scale of nonlocality we introduce in \S\S\ref{ssec:length-scales}.}. This is the main technical result of this paper and what is shown in \S\S \ref{sss:cnc}.

\subsection{A Planck Scale and a String Scale}\label{ssec:length-scales}

The holographic bulk of D$p$-brane systems has three essential length-scales\footnote{More precisely, two independent unitless ratios of length-scales.} -- the scale of the geometry (i.e. the AdS radius) $r_{\Lambda}$, the string scale $l_s \sim \sqrt{\alpha'}$, and the Planck length $l_p$. The string scale in particular is the scale of nonlocality -- the Ryu-Takayanagi formula picks up $\alpha'$ corrections as extremal surfaces become more and more curved \cite{Dong:2013qoa,Wall:2015raa}, and we expect that the area law contribution will cease to dominate below the string scale. In \cite{Maldacena:1997re,itzhaki1998supergravity}, it is emphasized that the ratios of these length-scales are given by
\begin{equation}\label{eqn:holo-length-scales}
\frac{l_s}{r_{\Lambda}} = \lambda^{-1/3} = \left(g_{\text{YM}}^2N\right)^{-1/3}, \quad \frac{l_p}{r_{\Lambda}} = N^{-1/4}.
\end{equation}
$\lambda$ is the `t Hooft coupling, and sets the string scale in the bulk. In the holographic dualities considered in \cite{Maldacena:1997re,itzhaki1998supergravity} and related works, the `t Hooft coupling (and therefore the ratio between the string scale and the AdS length-scale) is $O(1)$ in $N$ scaling. The Planck length $l_p$, meanwhile, is separated from the AdS lenghtscale and the string scale by nontrivial factors of $N$.

The tensor networks we consider naturally impose on us three length-scales that have a behavior strikingly similar to those in \eqref{eqn:holo-length-scales}. Consider Figs. \ref{fig:MTN-sketch-2} and \ref{fig:network-apds-intro}, and call the size of the geometry $\tilde{r}_{\Lambda}$. One natural length-scale to consider is the typical separation between nodes of the network, or equivalently between D0-branes in Fig. \ref{fig:brane-net}. This is analogous to $l_p$ -- it is the shortest possible distance scale, and sets the total number of degrees of freedom. The other length-scale is the typical lengths of the tensor network edges -- the purple $\lambda_{nn'}$ edges in the two figures. They need not only connect nearest-neighbor nodes, and may in principle stretch long distances along the geometry. This is the scale of nonlocality of the network, and behaves similarly to the string scale $l_s$. We refer to these scales as $\tilde{l}_p$ and $\tilde{l}_s$ respectively. The ratio of these two length-scales is set by the structure of $H_2$ in \eqref{eqn:full-ham-intro}, and in particular by the strength and structure of the couplings $g_i$ in the derivative expansion.

Note that in light of the D0-brane origin of holographic MQM (as in Fig. \ref{fig:brane-net}), the interpretation of this scale of nonlocality as a string scale might be taken quite literally. Below this scale, the off-diagonal matrix elements $x^i_{nn'}$ and $\lambda_{nn'}^{\alpha}$ modes connecting branes arehighly excited and entangled with nearby degrees of freedom. Above this scale, $x^i_{nn'}$ and $\lambda_{nn'}^{\alpha}$ are suppressed to their ground states. This exactly matches the behavior of off-diagonal matrix elements (i.e. string modes connecting branes) in D-brane matrix quantum mechanics \cite{taylor2001m,Polchinski:1999br}.

To find the nice property that area law entanglement is simply given by the number of tensor network legs crossing a given entanglement cut, we find in \S\ref{sec:matrix-tn} that we must consider the regime
\begin{equation}\label{eqn:string-ratio}
N^{-1/2} \ll \frac{\tilde{l}_s}{\tilde{r}_{\Lambda}} \ll 1.
\end{equation}
In particular, the `string scale' must be much greater than the `Planck scale'. It is perhaps an interesting coincidence that this regime includes the analog of the `t Hooft limit. The string scale of the matrix tensor networks we define depends on the Hamiltonian we choose, and we will find area law entanglement only above the string scale we set.

\subsection{Layout of the Paper}
In \textbf{\S\ref{sec:fuzzy-review}} we review the relevant main ideas of noncommutative geometry. We also argue that the matrices representing the noncommutative coordinate functions have a natural interpretation as the adjacency matrices of a graph.

\textbf{\S \ref{sec:matrix-tn}} contains the main results of the paper. We embed the structure of tensor networks into the MQM Hilbert space. We find that these noncommutative tensor networks have a natural scale of nonlocality, mimicking a string scale. This string scale may be tuned arbitarily small in comparison to the size of the geometry in the large $N$ limit, similar to the limit of strong `t Hooft coupling. We show that the low-energy states of reasonable $U(N)$-invariant MQM Hamiltonians produce states with area-law entanglement above this string scale, demonstrating that local geometry may emerge dynamically for these models.

In \textbf{\S \ref{sec:disc}} we discuss the results and comment on future directions.

\section{Fuzzy Coordinate Functions as Adjacency Matrices}\label{sec:fuzzy-review}

Noncommutative geometry has a storied history within string theory in general \cite{seiberg1999string,douglas2001noncommutative,szabo2003quantum,witten1986non,Ardalan:1998ce} and matrix quantum mechanics in particular \cite{taylor2001m,Steinacker_2010,Steinacker:2011ix,Han:2019wue}. The bulk of holographic systems, especially the compact dimensions, exhibit phenomena similar to physics on noncommutative backrounds \cite{McGreevy:2000cw,ho2000large,ho2001fuzzy}. In this section we briefly review how noncommutative manifolds are defined and how they naturally emerge in MQM systems.

Noncommutative manifolds (which we denote $\mathcal{M}_{\theta}$), much like their commutative counterparts, are defined through their algebra of functions and metric structure. The essential aspect is that the coordinate functions $x^i$ parameterizing $\mathcal{M}_{\theta}$ do not commute, and instead obey some relations
\begin{equation}
[x^i,x^j] = \theta^{ij}(x).
\end{equation}
The algebra of functions on $\mathcal{M}_{\theta}$ is then generated by all sums and products of the $x^i$. Perhaps surprisingly, from this basic starting point we may introduce a differential and integral structure. Consider some function $f(x)$ on $\mathcal{M}_{\theta}$. Its derivatives are given by
\begin{equation}
[x^i,f(x)] = \theta^{ij}\partial_j f(x).
\end{equation}
This definition is particularly natural because commutators obey the Leibniz rule. The Laplacian, which in turn determines the metric structure, is defined as
\begin{equation}
\Delta f(x) = \sum_i [x^i,[x^i,f(x)]].
\end{equation}
Integrals are given by traces --
\begin{equation}
\int_{\mathcal{M}_{\theta}} \sqrt{g}d^dx\, f(x^i) = \Tr[f].
\end{equation}

The simplest way to work with noncommutative manifolds is by finding a representation of their algebra. The prototypical example is the fuzzy sphere, which is defined by the relations
\begin{equation}\label{eqn:fuzz-sphere}
X^2 + Y^2 + Z^2 = R^2 \mathbb{1}, \quad [X^i,X^j] = i\epsilon^{ijk}X^k. 
\end{equation}
We recognize these expressions as the defining relations of $\mathfrak{su}(2)$ representations, and may find an explicit realization of the fuzzy sphere by choosing an irrep of $\mathfrak{su}(2)$ --
\begin{equation}
Z = \sum_{m=-N/2}^{N/2} |m)(m|, \quad X + iY = \sum_{m=-N/2}^{N/2-1}\sqrt{(N/2+m)(N/2-m+1)}|m)(m+1|.
\end{equation}
This is the basis in which $Z$ is diagonal and its eigenvalues are ordered. We interpret the spectrum of $Z$ as the distribution of $D0$-brane positions over the $z$ axis. That the eigenvalues are evenly spaced is a signature of the commutative sphere measure being $\sin \theta d\theta d\phi = d\phi dz$.

With an explicit representation given by $N \times N$ matrices, there is a close relationship between derivatives and length scales. In particular, consider some function on a noncommutative manifold represented by a matrix $F$ with entries $F_{nn'}$. Its commutator with a diagonal coordinate matrix (say $Z$) is
\begin{equation}\label{eqn:nonc-length-scale}
[Z,[Z,F]]_{nn'} = (z_n - z_n')^2F_{nn'}. 
\end{equation}
The ratio $\Tr[F^{\dag}[Z,[Z,F]]]/\Tr[F^{\dag}F]$ therefore measures the typical distance between $z$ coordinates that elements of $F$ stretch between, weighted by the size of the elements $F_{nn'}$. By extension, the noncommutative gradient $\Tr[F[X,[X,F]] + F[Y,[Y,F]] + F[Z,[Z,F]]]$ measures the average length squared $(x_n - x_{n'})^2 + (y_n - y_{n'})^2 + (z_n - z_{n'})^2$ of the matrix elements excited in $F$, and therefore sets a length-scale for $F$.

\begin{figure}[ht]
\centering
\includegraphics[width=0.8\textwidth]{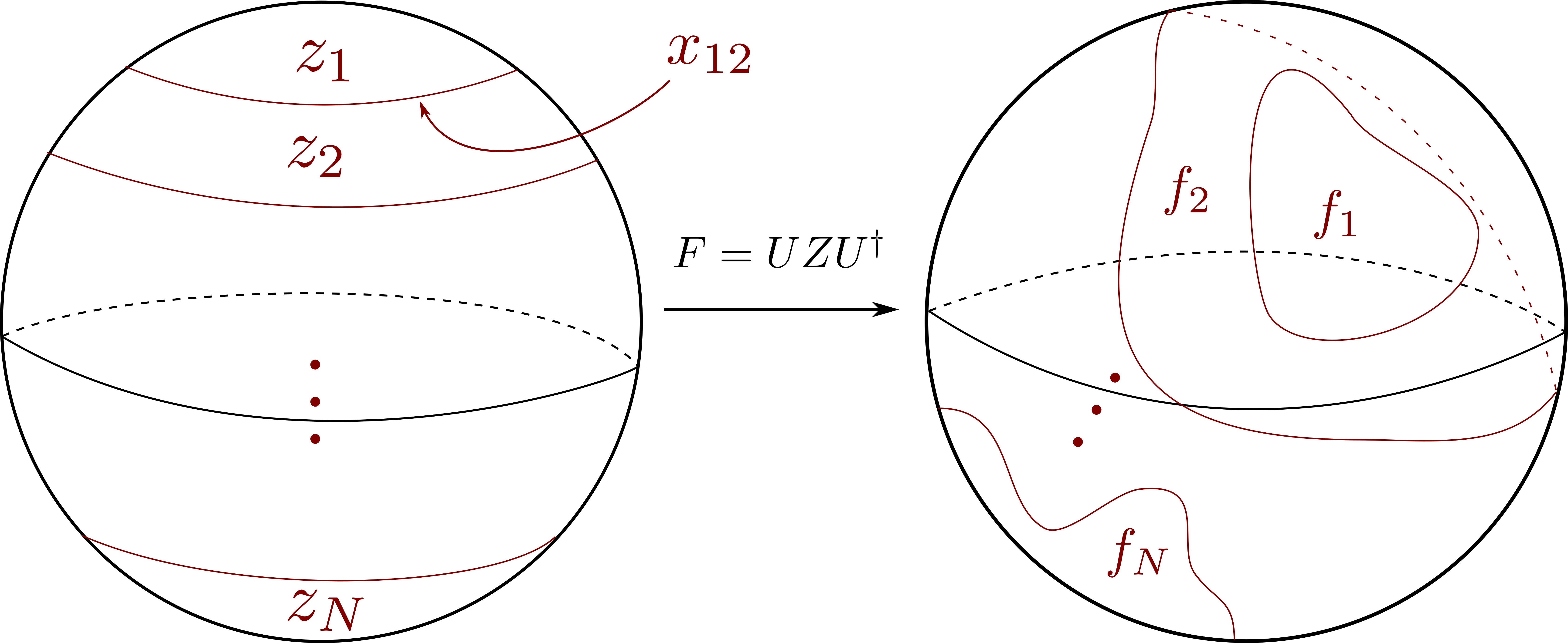}
\caption{The $Z$ basis, pictured on the left, is one in which each $D0$-brane on the fuzzy sphere has a well-defined $Z$ coordinate. We can transform to another basis by diagonalizing a different Hermitian matrix $F$. We interpret $F$ as a different curvilinear coordinate $f(x^i)$, and the basis transformation as taking us to a new coordinate system where each $D0$ brane has a well defined $f$ coordinate. This coordinate change is interpreted as the fuzzy regularization of an area-preserving diffeomorphism (as in \cite{hoppe1982quantum,swain2004topology}). In each basis, the coordinate matrices $X^i$ only have nonzero elements (to leading order in $1/N$) if the $n$th and $n'$th $D0$-branes have adjacent $f$ coordinates (i.e. share a boundary on the right hand figure).}
\label{fig:sphere-coords}
\end{figure}

In the basis where $Z$ is diagonal, $X$ and $Y$ are sparse matrices with support only on the diagonals adjacent to the main diagonal. This means that e.g. in Fig. \ref{fig:sphere-coords}, $X_{nn'}$ is only nonzero in a particular $U(N)$ basis if $n$ and $n'$ are adjacent D0 branes in the coordinate system corresponding to this basis. In this sense, $X$ and $Y$ precisely behave as adjacency matrices of a graph. Digging a bit deeper, $|x_{nn'}|$ is precisely the length of the perimeter between the $n$th and $n'$th D0-branes. The reason for this is that projection matrices $\Theta_{\Sigma}$ are the noncommutative analog of step functions, and their derivatives $[X^i,\Theta_{\Sigma}]$ (the natural noncommutative analog of a delta function) pick out the off-diagonal blocks of the coordinate matrices $X^i$.

So far we have taken the coordinate functions $X^i$ to be fixed. One way in which noncommutative geometry naturally arises in MQM is by interpreting the dynamical Hermitian matrices in \eqref{eqn:mqm-lag} as the coordinate matrices of a noncommutative manifold $\mathcal{M}_{\theta}$. Because the metric structure is determined by the $X^i$, if we make the fuzzy coordinates dynamical we also make the metric dynamical. The fuzzy sphere, for example, arises as one possible vacuum state in the BMN (or mini-BMN \cite{Han:2019wue,Anous:2017mwr}) matrix model. \eqref{eqn:mqm-lag} is precisely designed to energetically suppress string modes propagating between distant D-branes, and we can see this effect explicitly from \eqref{eqn:fuzz-sphere} in that only the matrix elements between adjacent D-branes are populated with nonzero entries. In this way, it is natural for low-energy states of MQM models to behave as dynamical adjacency matrices of local graphs that only have connections between nearby nodes. Because the geometry and topology of these graphs are determined by the dynamics of the theory, they come with a manifest sense of background independence.

Intuition tells us that in low energy states of quantum systems degrees of freedom which are more strongly coupled are more entangled. Moreover, it is often tempting to interpret strings stretching across an entanglement cut as extended degrees of freedom which should be cut to produce entanglement edge modes \cite{Susskind:1994sm,levin2006detecting,ghosh2015entanglement,hampapura2021target, Frenkel:2021yql, frenkel2023emergent}. The way in which we make this intuition precise is by interpreting the graph arising from the adjacency matrix picture of $X^i$ as the graph of a tensor network. This is the task we take up in \S\ref{sec:matrix-tn}.

\subsection{Subregions of Noncommutative Manifolds}\label{ssec:noncomm-sub}

Considering subalgebras and entanglement on the tensor networks we are defining is closely related to the idea of considering geometric subregions of noncommutative geometries. On a typical commutative manifold, a subregion $\Sigma$ is defined via a characteristic function $\chi_{\Sigma}(x^i)$. $\chi_{\Sigma}(x^i)$ evaluates to 1 inside $\Sigma$ and evaluates to 0 on the complement $\bar{\Sigma}$.  Any function $f(x^i)$ may be decomposed into two parts:
\begin{equation}
f_{\Sigma}(x^i) := \chi_{\Sigma}(x^i)f(x^i), \quad f_{\bar{\Sigma}}(x^i) := (1 - \chi_{\Sigma}(x^i))f(x^i).
\end{equation}

\begin{figure}[ht]
\centering
\includegraphics[width=0.7\textwidth]{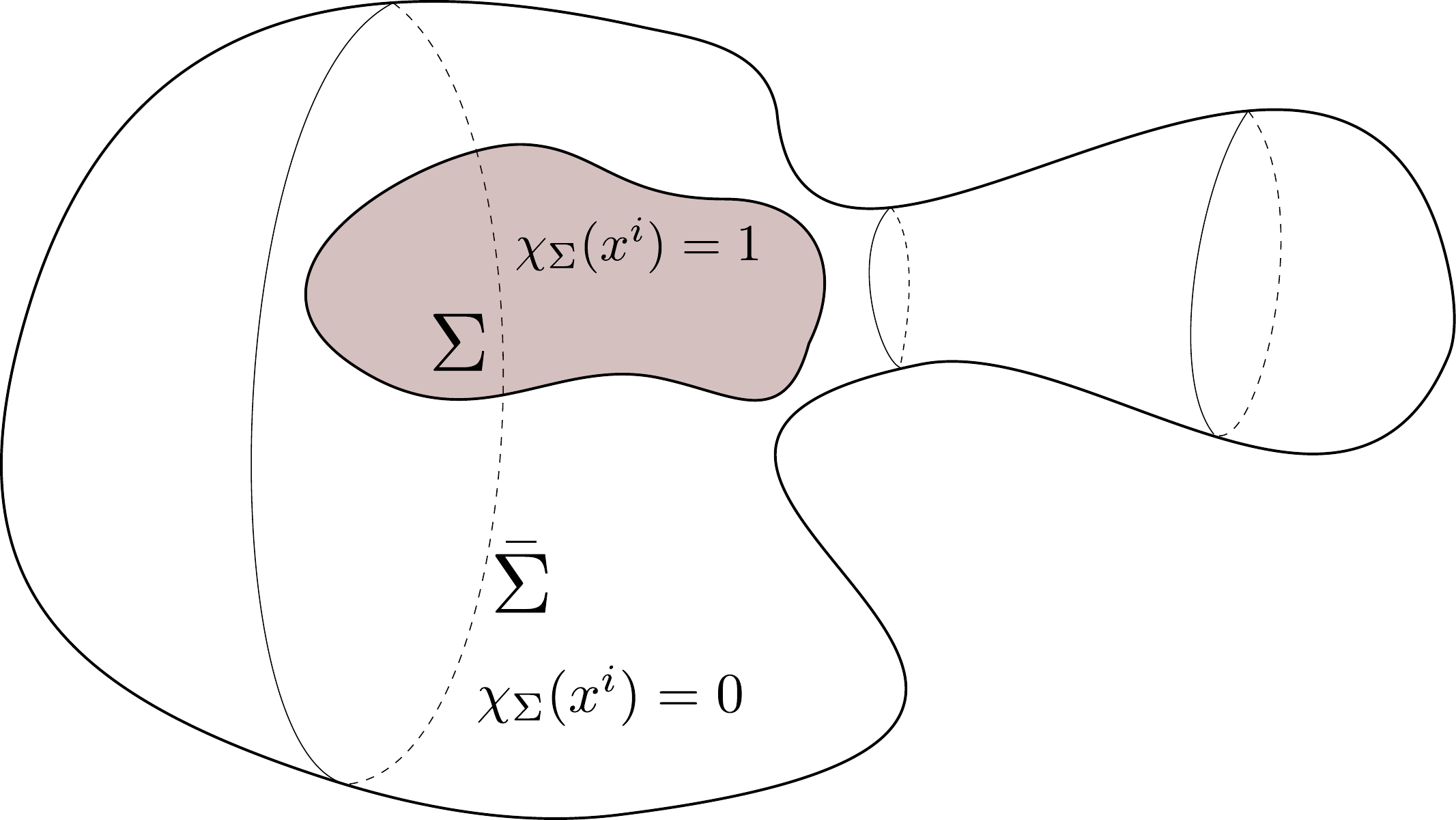}
\caption{A subregion $\Sigma$ of a compact commutative manifold. The characteristic function $\chi_{\Sigma}(x^i)$ evaluates to 1 inside $\Sigma$ and evaluates to 0 outside of $\Sigma$. When we consider noncommutative manifolds, $\chi_{\Sigma}(x^i)$ will be promoted to a projection matrix $\Theta_{\Sigma}$ and all functions $F$ will decompose into four distinct pieces (as in Fig. \ref{fig:MTN-subblocks}.) For an analogous decomposition for the graph structure we described in the previous section, see Fig. \ref{fig:graph-partition}.}\label{fig:manifold-subregion}
\end{figure}

\begin{figure}[ht]
\centering
\includegraphics[width=0.45\textwidth]{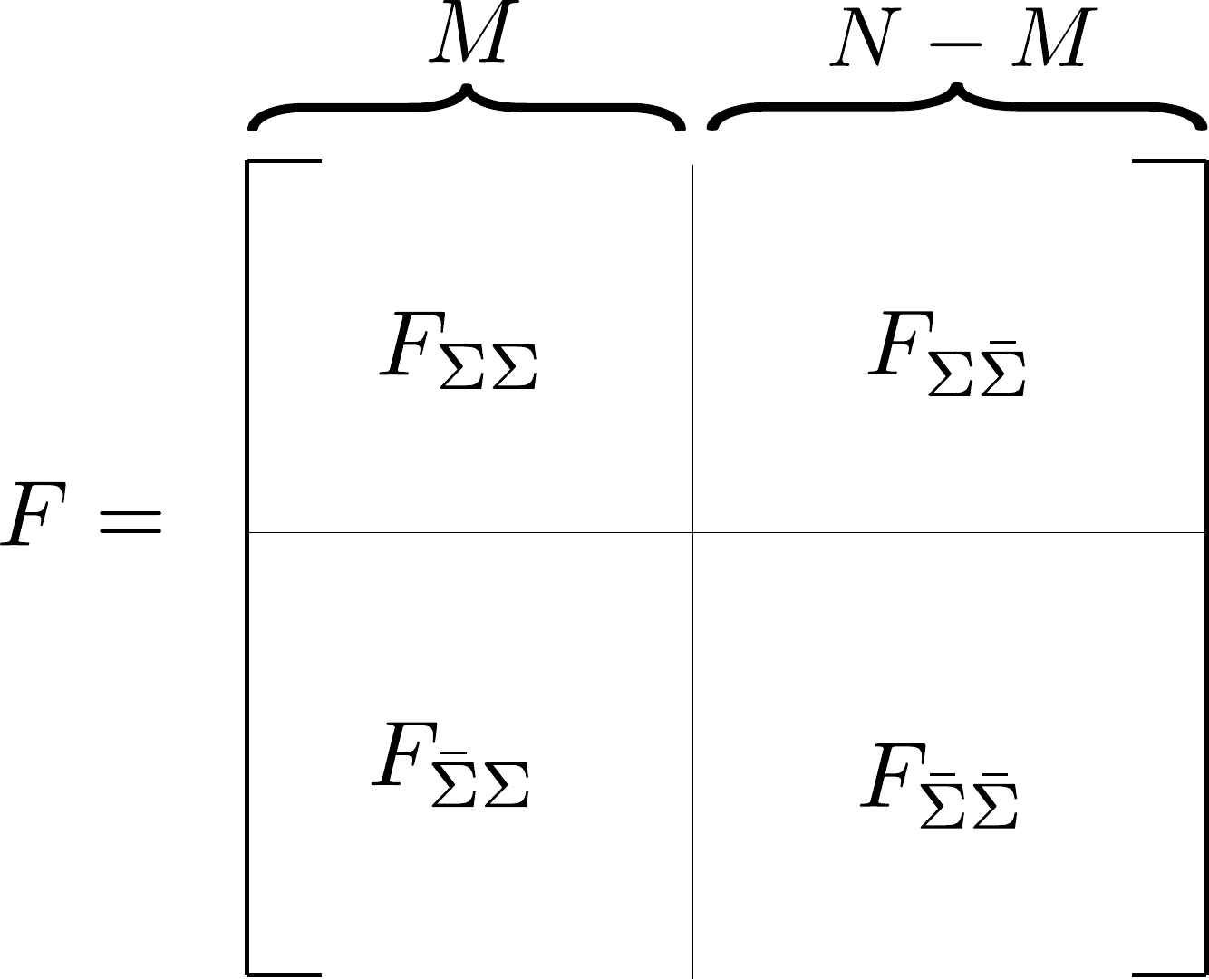}
\caption{A depiction of the decomposition of a function on a noncommutative manifold. As depicted in Fig. \ref{fig:graph-partition}, $F_{\Sigma \Sigma}$ is the piece of the function completely contained in the interior of $\Sigma$, $F_{\bar{\Sigma}\bar{\Sigma}}$ is the piece in the interior of $\bar{\Sigma}$, and the $F_{\bar{\Sigma}\Sigma}$, $F_{\Sigma \bar{\Sigma}}$ pieces are those that are in contact or cross the boundary of $\Sigma$ and may interact with modes in both regions.}\label{fig:MTN-subblocks}
\end{figure}

\begin{figure}[ht]
\centering
\includegraphics[width=0.9\textwidth]{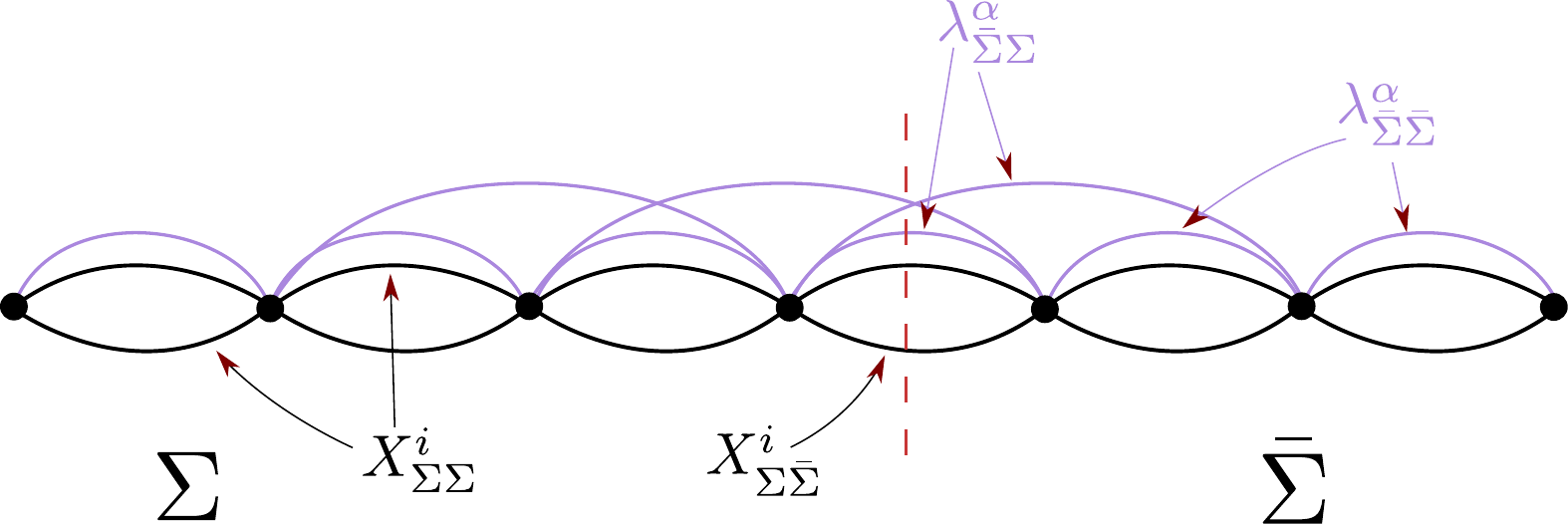}
\caption{A depiction, in terms of the graph language of Figs. \ref{fig:MTN-sketch}-\ref{fig:network-apds-intro}, of the structure of subregions of a noncommutative manifold. This is one way to picture the noncommutative analog of Fig. \ref{fig:manifold-subregion}. The block decomposition of Fig. \ref{fig:MTN-subblocks} should be kept in mind. The vertical dashed line is again the entanglement cut separating subregion $\Sigma$ on the left and subregion $\bar{\Sigma}$ on the right. $X^i_{\Sigma \Sigma}$, as an example, contains all the black geometric graph edges entirely contained within the region $\Sigma$. Similarly, $\lambda^{\alpha}_{\bar{\Sigma}\bar{\Sigma}}$ consists of all of the purple tensor network edges that are entirely contained in $\bar{\Sigma}$. Any matrix subblock with mixed indices such as $X^i_{\sigma \bar{\Sigma}}$ or $\lambda^{\alpha}_{\Sigma \bar{\Sigma}}$ consists of edges that cross the entanglement cut.}\label{fig:graph-partition}
\end{figure}

On a noncommutative manifold characteristic functions are promoted to projection matrices $\Theta_{\Sigma}$ satisfying $\Theta_{\Sigma}^2 = \Theta_{\Sigma}$. Unlike the commutative case, a noncommutative function $F$ has four distinct pieces to consider:
\begin{equation}
\begin{split}
&F_{\Sigma \Sigma} := \Theta_{\Sigma} F \Theta_{\Sigma}, \quad F_{\Sigma \bar{\Sigma}} := \Theta_{\Sigma} F (1 - \Theta_{\Sigma}),\\
&F_{\bar{\Sigma} \Sigma} := (1-\Theta_{\Sigma})F\Theta_{\Sigma}, \quad F_{\bar{\Sigma} \bar{\Sigma}} := (1-\Theta_{\Sigma})F (1 - \Theta_{\Sigma}).
\end{split}
\end{equation}
If we choose a basis where $\Theta_{\Sigma}$ is diagonal with ordered eigenvalues, this is nothing more than a block decomposition of the $N \times N$ matrix $F$ (see Fig \ref{fig:MTN-subblocks}). The blocks $F_{\Sigma \bar{\Sigma}}$ and $F_{\bar{\Sigma}\Sigma}$ are associated to the boundary of the noncommutative manifold, and in particular should be thought of as modes that cross the boundary (see Fig. \ref{fig:graph-partition}).

To consider RT surfaces, instead of minimizing over a discrete set of entanglement cuts we must minimize over the continuous set of projection matrices. This type of minimization will appear shortly in \cite{ffhs:2024xx}, and would be interesting to directly apply to entropy calculations and bulk reconstruction in the tensor networks we are considering.

\section{Embedding Tensor Networks into MQM}\label{sec:matrix-tn}

\subsection{The Embedding Map}\label{sec:emb-map}

We interpret the color indices $n$ of the matrices as nodes in the network. To construct quantum states, we introduce dynamical degrees of freedom in the form of $N_f$ matrices of spinors $\lambda^{a}$ and $N_f$ vectors of spinors $\psi^a$. For simplicity, instead of treating these spins as fermions we take their (anti-)commutation relations to be
\begin{equation}\label{eqn:MQM-alg}
\begin{split}
&\{\lambda_{nn'}, (\lambda_{nn'})^{\dag}\} = \{\psi_n, \psi_n^{\dag}\} = 1\\
&[\lambda_{ab},\lambda_{cd}] = 0 \text{ if $a \neq c$ or $b \neq d$}, \quad [\psi_a, \psi_b] = 0 \text{ if $a \neq b$}.
\end{split}
\end{equation}
We take this unusual choice to simplify the representations of the operators, so that we may take e.g.
\begin{equation}
\lambda_{nn'} = \begin{bmatrix}
0& 1\\
0 & 0
\end{bmatrix} \otimes \bigotimes_{(n_1,n_2)\neq (n,n')}\mathbb{1}_{nn'} \quad \lambda_{nn'}^{\dag} = \begin{bmatrix}
0& 0\\
1 & 0
\end{bmatrix} \otimes \bigotimes_{(n_1,n_2)\neq (n,n')}\mathbb{1}_{nn'}.
\end{equation}
With these conventions, the generators of $U(N)$ transformations on the $X$, $\lambda$, $\psi$ variables take the form
\begin{equation}\label{eqn:U-gens}
\begin{split}
&G^X_{nn'} = 2i\sum_{n''}\left(X_{nn'}\Pi_{n'n''} - X_{n''n'}\Pi_{nn'}\right),\\
&G^{\lambda}_{nn'} = 2i\sum_{n''}\left(\lambda_{nn''}\lambda^{\dag}_{n''n'} - \lambda^{\dag}_{nn''}\lambda_{n''n'}\right) - 4i\delta_{n,n'} \lambda_{nn}^{\dag}\lambda_{nn},\\
&G^{\psi}_{nn'} = \psi^{\dag}_{n'} \psi_n.
\end{split}
\end{equation}

The $N_f$ creation operators $\lambda_{nn'}^{a \dag}$ act on the $D:=2^{N_f}$ dimensional Hilbert space $\mathbb{H}_{nn'}$ that lives on the link connecting nodes $n$ and $n'$. On each factor of the Hilbert space, we label as $\ket{0}$ the states annihilated by $\lambda_{nn'}$ and $\ket{1}$ the states annihilated by $\lambda^{\dag}_{nn'}$. We interpret the dynamical variables $\psi_n$ as living on `external' legs of the tensor network with corresponding spaces $\mathbb{H}_n$. For holographic models these can be interpreted as either boundary or bulk external legs, depending on where the node $n$ is located in the fuzzy geometry. For notational convenience we introduce a basis of states $\ket{\mu}_{nn'}$ or $\ket{\mu}_n$ on the corresponding Hilbert spaces, with the index $\mu$ running from 0 to $D-1$.

\begin{figure}[ht]
\centering
\includegraphics[width=0.5\textwidth]{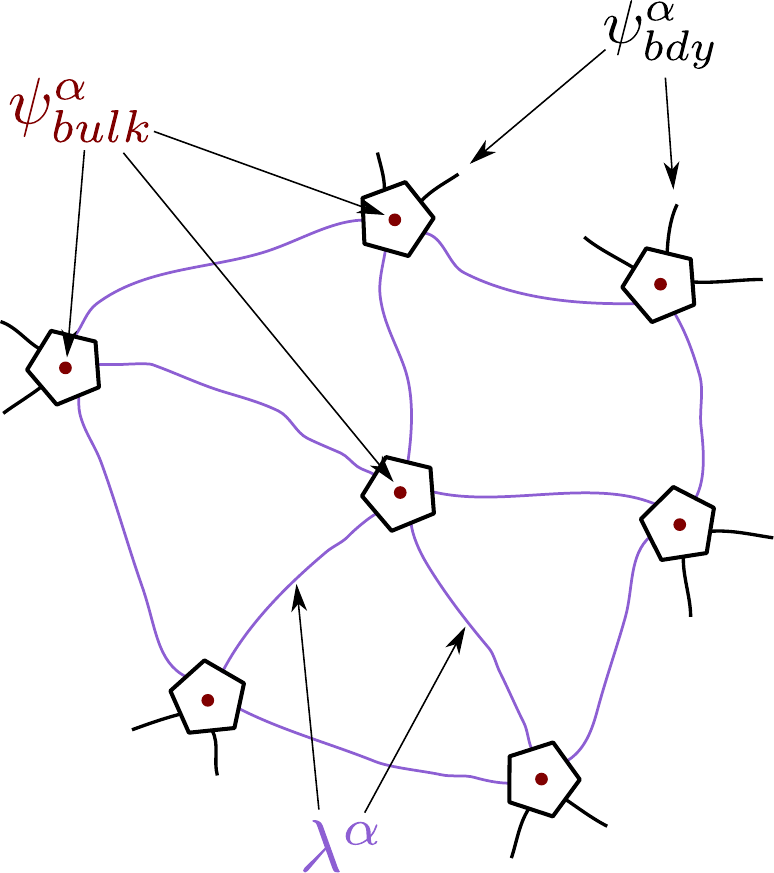}
\caption{A depiction of how the MQM degrees of freedom in \eqref{eqn:MQM-alg} map onto a tensor network as in \cite{pastawski2015holographic} and Fig. \ref{fig:tensor-net}. $\lambda^{\alpha}$ set the state of internal legs, and $\psi^{\alpha}$ the state of external legs (either bulk or boundary).}\label{fig:tensor-net-labeled}
\end{figure}

We have now introduced all the degrees of freedom necessary to construct any tensor network state depending on what state $\ket{\chi}_{nn'}$ we fix on the products $\mathbb{H}_{nn'} \otimes \mathbb{H}_{n'n}$. For the most straightforward type of network, we can create a link (or a lack of a link) between the nodes $n$ and $n'$ by choosing one of two states on these spaces:
\begin{equation}\label{eqn:Hilbert-space}
\begin{split}
\ket{\chi}_{nn'} &= \frac{1}{\sqrt{D}}\sum_{\mu}\ket{\mu}_{nn'} \otimes \ket{\mu}_{n'n} \quad \Longleftrightarrow \quad \text{(there \textbf{is} a link between nodes $n$ and $n'$)},\\
\ket{\chi}_{nn'} &= \ket{0}_{nn'} \otimes \ket{0}_{n'n} \quad \Longleftrightarrow \quad \text{(there \textbf{is not} a link between nodes $n$ and $n'$)}.
\end{split}
\end{equation}
Of course, as \cite{Hayden:2016cfa,Cheng:2022ori} point out, it is fruitful to consider more general types of connections between nodes besides maximally entangled states. With this in mind, we generally take a state in $\mathbb{H}_{\lambda}$ as defining the graph the tensor network lives on, so we denote elements of $\mathbb{H}_{\lambda}$ as $\ket{\Gamma}$.

To introduce random tensor networks, random tensors are defined on the states
\begin{equation}
\ket{V_n} \in \tilde{\mathbb{H}}_n := \mathbb{H}_n \otimes \bigotimes_{n'} \mathbb{H}_{nn'}, \quad \dim \tilde{\mathbb{H}}_n = D^{N + 1} = 2^{(N + 1)N_f}.
\end{equation}
Correspondingly, given a `graph state' $\ket{\Gamma}$ the state on the external links is given by
\begin{equation}
\ket{\Psi} = \bra{\Gamma} \bigotimes_{n} \ket{V_n}.
\end{equation}

\subsection{Partitions and Subalgebras}

Now that we have described the Hilbert space of the theory \eqref{eqn:Hilbert-space}, we explain how to consider subregions of MQM systems and matrix tensor networks in particular. This story is essentially the same as that of target space entanglement in MQM \cite{das2020bulk,Das:2020jhy,karczmarek2014entanglement,hampapura2021target,Han:2019wue,frenkel2023emergent,Frenkel:2021yql}, but contains a crucial unique ingredient in how we treat the off-diagonal matrix elements.

\begin{figure}[ht]
\centering
\includegraphics[width=0.7\textwidth]{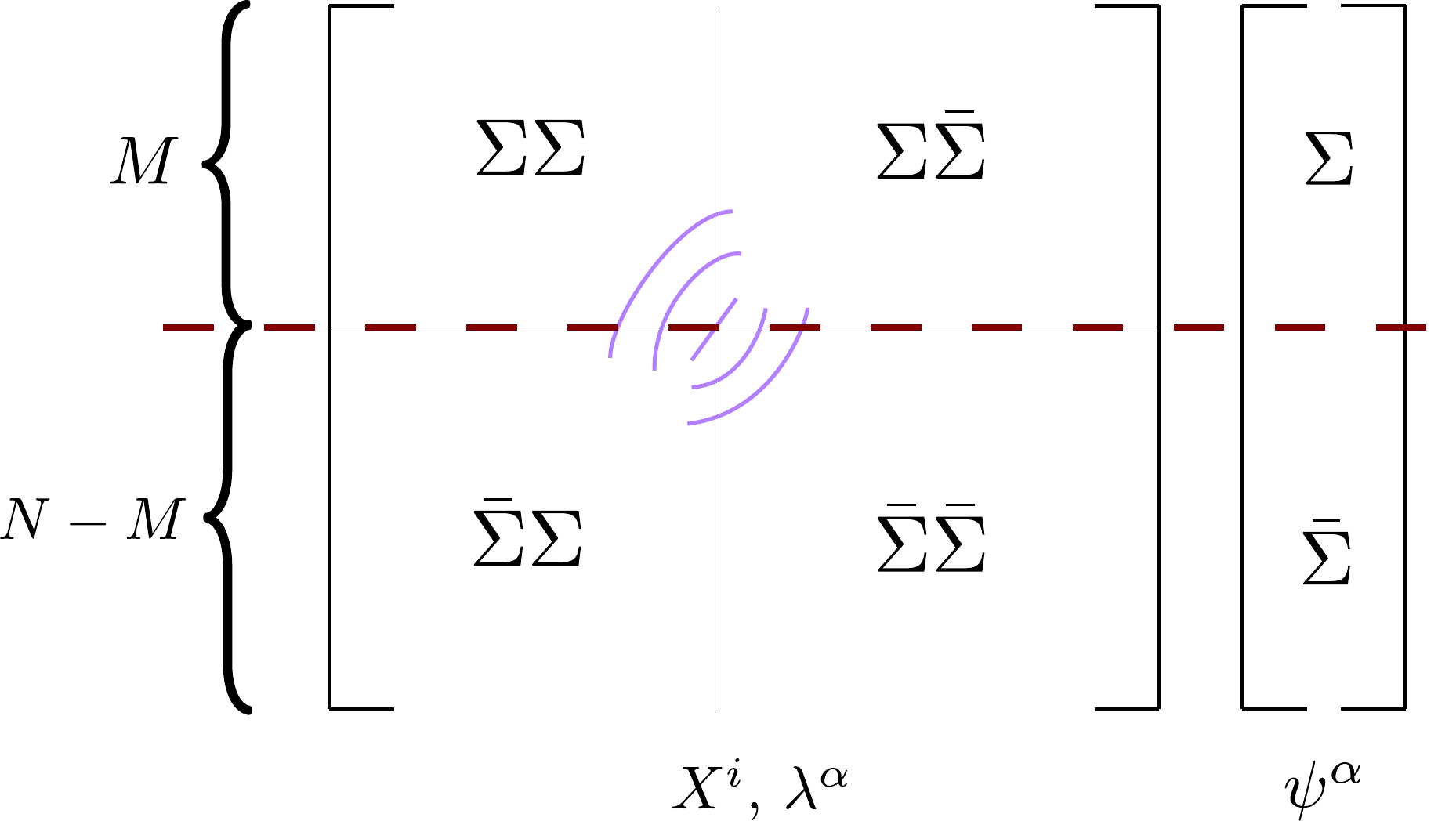}
\caption{A schematic of how to partition the degrees of freedom of a matrix tensor network. We write the state in a particular $U(N)$ frame, then take the top $M$ rows of all matrix degrees of freedom $X^i$ and $\lambda^{\alpha}$. The entanglement cut is represented by the dashed line across the figure. The dominant contribution to the entanglement entropy is the entanglement of qubits in the $\Sigma \bar{\Sigma}$ block with those in the $\bar{\Sigma}\Sigma$ block. To emphasize this, the purple lines connecting the $\Sigma \bar{\Sigma}$ and $\bar{\Sigma} \Sigma$ blocks are the same as the edges of the tensor network diagram in Figs. \ref{fig:MTN-sketch-2} and \ref{fig:graph-partition} that are cut by the entanglement cut.}\label{fig:MTN-partition}
\end{figure}

Remember that each matrix entry $nn'$ has a Hilbert space associated to it. This is true of both the $X^i$ and the $\lambda^{\alpha}$. The essential idea, then, is to choose an appropriate basis in which to write our vectors $\psi^{\alpha}$ and matrices\footnote{This should be thought of as a gauge fixing condition for our $U(N)$ invariant wavefunctions.} and split the system along the row index (see Fig. \ref{fig:MTN-partition}). More precisely, we write
\begin{equation}
\mathbb{H} = \mathbb{H}_{\Sigma} \otimes \mathbb{H}_{\bar{\Sigma}}, \quad \mathbb{H}_{\Sigma} = \bigotimes_{n \leq M} \mathbb{H}_n \otimes \bigotimes_{n'}\mathbb{H}_{nn'}, \quad \mathbb{H}_{\bar{\Sigma}} = \bigotimes_{n > M} \mathbb{H}_n \otimes \bigotimes_{n'}\mathbb{H}_{nn'}.
\end{equation}

The most convenient way to split a matrix is to introduce a projection matrix $\Theta_{\Sigma}$ on color space. The label $\Sigma$ may be associated to a subregion of the noncommutative manfiold of area $M := \Tr[\Theta_{\Sigma}]$ (see \S\S\ref{ssec:noncomm-sub}). $\Theta_{\Sigma}$ is the noncommutative analog of a characteristic function, which evaluates to 1 on some subset $\Sigma$ of a manifold and evaluates to 0 on the complement $\bar{\Sigma}$ of $\Sigma$.

\subsection{Dynamically Introducing Locality}

Thus we have introduced the tools to in principle describe tensor networks on potentially all-to-all connected (or otherwise nonlocal) graphs. To make contact with the typical notion of tensor network, however, we want there to be a sense in which only nearby nodes are connected. We could by hand choose a graph state $\ket{\Gamma} \in \mathbb{H}_{\lambda}$ that reproduces a local network, but this simply reduces us to the construction explored in \cite{Hayden:2016cfa}. In this section consider some simple MQM-like Hamiltonians and argue that their low-energy states have the properties of tensor networks with only geometrically nearby nodes connected.

We consider two classes of MQM models. The first are based on Matrix Quantum Hall models \cite{Susskind:2001fb,Hellerman:2001rj,Polychronakos:2001mi,Tong:2015xaa} where there are two coordinate matrices $X$ and $Y$ that commute at a classical level but not as quantum operators. This is the usual way by which noncommutative geometry arises in quantum hall models, as $x$ and $y$ become conjugate phase space variables in the limit of a strong magnetic field. We refer to this setup as \textbf{classically commutative, quantum noncommutative}, and explore it in \S\ref{sss:cc-qnc}. This is the notion of noncommutative geometry considered in e.g. \cite{Zhu:2022gjc}.

The second class consists of field theories on honest noncommutative geometries with classically noncommuting coordinates. We refer to this as the \textbf{classically noncommutative} setup and discuss it in \S\ref{sss:cnc}. This setup has a much richer structure and is closer in spirit to MQM with multiple matrices.

\subsubsection{Classically commutative, Quantum noncommutative geometries}\label{sss:cc-qnc}

\begin{figure}[ht]
    \centering
    \includegraphics[width=0.65\textwidth]{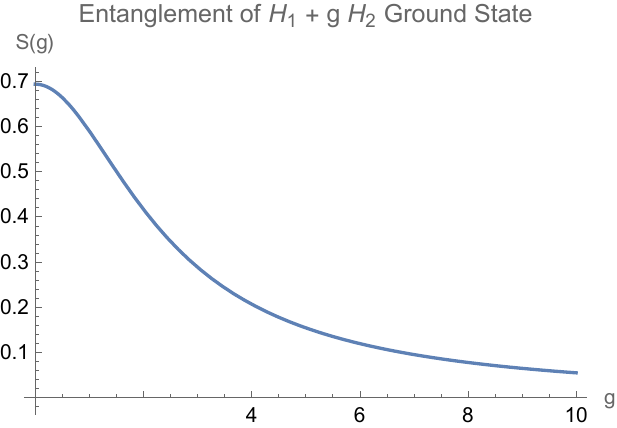}
    \caption{A plot of \eqref{eqn:qubit-ent} -- the entanglement of the ground state of the Hamiltonian $H_1 + gH_2$ as a function of $g$ as in \eqref{eqn:cnc-Ham}. For large $g$, this function falls off as $S(g) \approx \log(eg^2)/g^2$.}\label{fig:g-entanglement}
\end{figure}

As in \cite{Polychronakos:2001mi,Tong:2015xaa} we consider a complex matrix $Z = X + iY$. We make the assumption that $[Z,Z^{\dag}] = 0$, so that $Z$ is unitarily diagonalizable. We first consider the case where the matrix $Z$ is fixed and has eigenvalues $z_i$. For simplicity, we drop the $\alpha$ index on the $\lambda$ and $\psi$ spinors for now. The Hamiltonian we consider is
\begin{equation}
H = \Tr[i(\lambda^{\dag 2} - \lambda^2) + \lambda^{\dag}[Z,[Z^{\dag}, \lambda]]].
\end{equation}
We choose to work in a basis where $Z$ is diagonal. In components, this Hamiltonian is written as
\begin{equation}\label{eqn:cnc-Ham}
\begin{split}
&H = \sum_{nn'}H_{nn'}, \quad H_{nn'}:=\left(H_{1,nn'} + |z_n - z_{n'}|^2H_{2,nn'}\right)\\
&H_{1,nn'} := 2i\sum_{nn'}\left(\lambda_{nn'}^{\dag}\lambda^{\dag}_{n'n} - \lambda_{nn'}\lambda_{n'n}\right), \quad H_{2,nn'} := 2\sum_{nn'}\lambda_{nn'}^{\dag}\lambda_{nn'}
\end{split}
\end{equation}
This Hamiltonian decouples on each Hilbert space factor $\mathbb{H}_{nn'} \otimes \mathbb{H}_{n'n}$, where each $\mathbb{H}_{n'n}$ is two dimensional. The ground state of $H_{1,nn'}$ is the maximally entangled state $\frac{1}{\sqrt{2}}\left(\ket{11} + \ket{00}\right)$. The ground state of $H_{2,nn'}$ is $\ket{00}$. The entanglement $S$ of the ground state of $H_{1,nn'} + gH_{2,nn'}$ as a function of the coupling $g$ is
\begin{equation}\label{eqn:qubit-ent}
S(g)=-\frac{1}{4}\left|\sqrt{4 + g^2}-g\right|^2\log(\frac{1}{4}\left|\sqrt{4 + g^2}-g\right|^2).
\end{equation}
Its behavior is shown in Fig. \ref{fig:g-entanglement}, but the only important point about its functional form is that it is $O(1)$ for $g < 1$ and falls off as $\log(eg^2)/g^2$ for $g \gg 1$. The overall ground state wavefunction will simply be a tensor product of the ground states of each $H_{nn'}$.

So far we have not moved beyond familiar tensor network graphs. To do so, we introduce dynamics on $Z$. We take the Hamiltonian studied in \cite{Polychronakos:2001mi,Tong:2015xaa}, which we refer to as the matrix quantum hall (MQH) model.
\begin{equation}
H = \Tr[Z^{\dag}Z + 2i \left(\lambda^{\dag 2} - \lambda^2\right) + \lambda^{\dag}[Z^{\dag},[Z,\lambda]]].
\end{equation}
The matrix elements of $Z$ obey the commutation relations
\begin{equation}
[Z_{ab},(Z^{\dag})_{cd}] = \delta_{ad}\delta_{bc},
\end{equation}
so without the coupling to $\lambda$ the system decomposes into $N^2$ variables obeying the harmonic oscillator algebra. 

We treat the system by passing to a collective field description, where we define the eigenvalue density
\begin{equation}
\rho(z) = \sum_n \delta(z - z_n). 
\end{equation}
In terms of $\rho$ the Hamiltonian takes the form\footnote{I would like to thank Laurens Lootens for pointing out the similarity to continuous tensor network models as in \cite{Tilloy:2018gvo}. One may view this construction as a fuzzy-regulated version of a continuous tensor network.}
\begin{equation}
\begin{split}
H = &\int \rho(z) d^2z\, |z|^2 +\\
&+2i\kappa \int \rho(z) \rho(z') d^2zd^2z'\,\left[\lambda^{\dag}(z,z')\lambda^{\dag}(z_2,z_1) - \lambda(z_1,z_2)\lambda(z,z')\right]+\\
&+ \kappa \int \rho(z_1) \rho(z_2) d^2zd^2z'\, |z - z'|^2\lambda^{\dag}(z,z')\lambda(z,z').
\end{split}
\end{equation}
We have introduced an additional coupling constant $\kappa$, which we may take to be small to ensure the bilocal fields $\lambda(z,z')$ do not strongly backreact on the geometry. Each $\lambda(z,z')$ mode now decouples, and the entanglement of a subregion $\Sigma$ with its complement $\bar{\Sigma}$ is now calculated to leading order in $N$ as
\begin{equation}\label{eqn:cnc-s-int}
S_{\Sigma} = \int_{\Sigma}d^2z \int_{\bar{\Sigma}}d^2z' \rho(z)\rho(z') s(|z-z'|).
\end{equation}
Noting that we may plug into \eqref{eqn:qubit-ent} $g = |z - z'|^2$, $s(|z-z'|)$ falls off as
\begin{equation}
s(|z-z'|) \approx \frac{\log (e |z-z'|^4)}{|z-z'|^4}\, , \quad |z-z'| \gg 1.
\end{equation}
With this in mind, the integral in \eqref{eqn:cnc-s-int} is dominated by the UV contribution localized to the entanglement cut, so we find an area law entanglement. 

\subsubsection{Classically noncommutative models}\label{sss:cnc}

We turn to a case where the background geometry on which the tensor network lives is honestly noncommutative at a classical level, as in \S \ref{sec:fuzzy-review}. For clarity and concreteness we focus on the fuzzy sphere as in \eqref{eqn:fuzz-sphere} \cite{Madore:1991bw}, but the results of this section are readily extended to any weakly curved noncommutative geometry with small noncommutativity parameter.

We first consider the case where the background geometry is fixed, so the $X^i$ don't fluctuate. The Hamiltonian on the $\lambda$ degrees of freedom takes the form as in \eqref{eqn:full-ham-intro}
\begin{equation}\label{eqn:full-ham}
\begin{split}
&H = H_1 + H_2\\
&H_1 = i\sum_{\alpha}\Tr[(\lambda^{\dag})^2 - \lambda^2]\\
&H_2 = \sum_{\alpha} \Tr[\lambda^{\dag}(g_1[X^i,[X^i,\lambda]] + g_2[X^i,[X^i,[X^j,[X^j,\lambda]]]] + \ldots)].
\end{split}
\end{equation}
Sums over the repeated $i,j$ indices are implicit. As in \S\S\ref{sss:cc-qnc}, we have dropped the $\alpha$ index as it just comes along for the ride. The ground state of \eqref{eqn:full-ham} is again easy to determine, this time by decomposing $\lambda$ into matrix spherical harmonics \cite{Taylor:1999qk,Han:2019wue}:
\begin{equation}\label{eqn:msh-ham}
\begin{split}
&\lambda = \sum_{jm}\lambda_{jm}\hat{Y}_{jm},\\
&H = \sum_{j,m>0}\lambda^{\dag}_{j,m}\lambda^{\dag}_{j,-m} - \lambda_{j,m}\lambda_{j,-m} + \sum_{jm}\epsilon(j)\lambda_{jm}^{\dag}\lambda_{jm},\\
&\epsilon(j) = \sum_k g_k j^k(j + 1)^k.
\end{split}
\end{equation}
The entanglement between the $\hat{Y}_{jm}$ mode and the $\hat{Y}_{j,-m}$ mode (for $m \geq 1$) is obtained by plugging $\epsilon(j)$ in the place of $g$ in \eqref{eqn:qubit-ent}.

\begin{figure}[ht]
\centering
\includegraphics[width=0.9\textwidth]{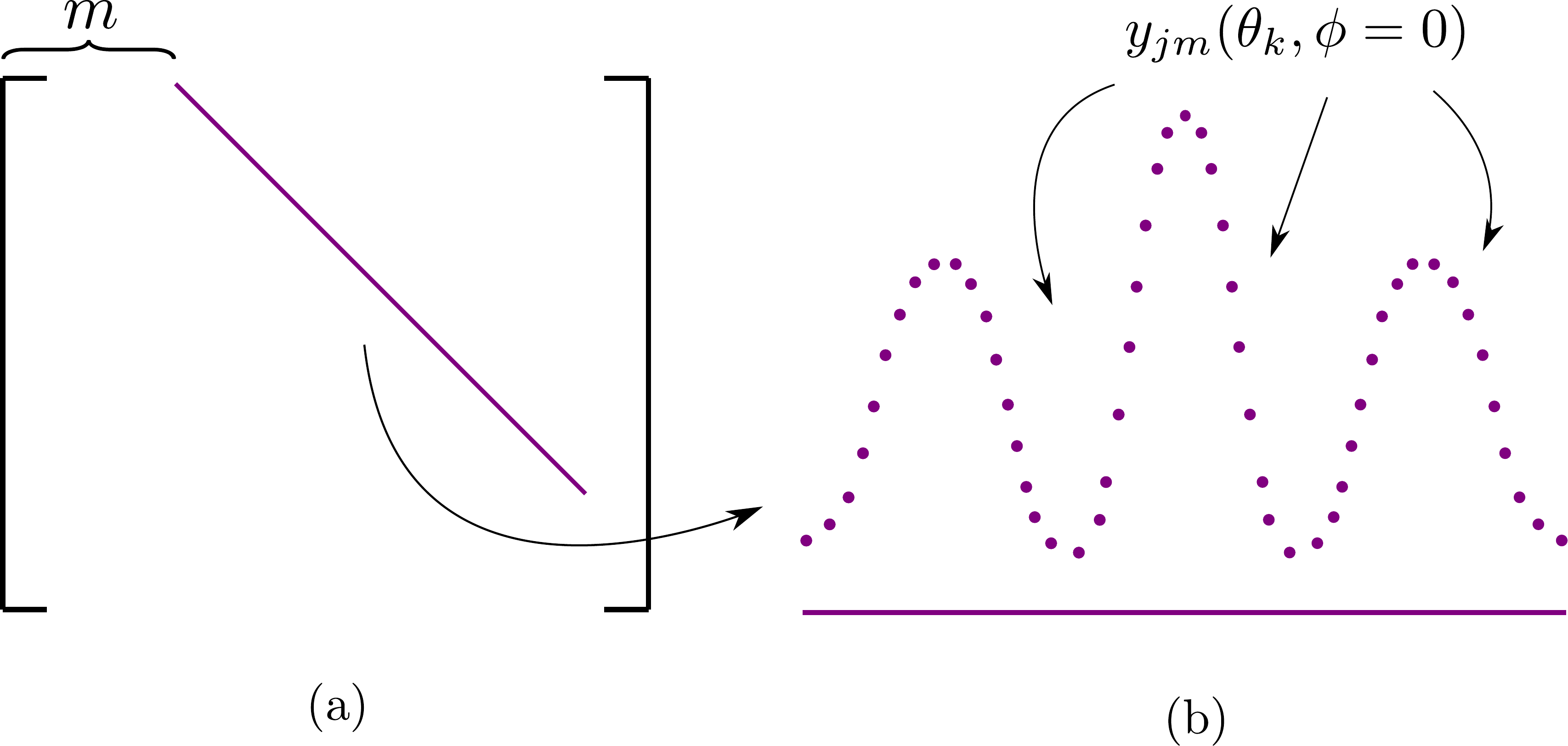}
\caption{We demonstrate the structure of matrix spherical harmonics. In the diagonal $Z$ basis, the matrix spherical harmonic $\hat{Y}_{jm}$ is supported only on the $m$th diagonal (as depicted in (a)). In (b) we depict the structure of the matrix entries themselves along this diagonal. They form a discretized version of the continuum spherical harmonics $y_{jm}(\theta,\phi)$, evaluted at $\phi = 0$ and the appropriate points $\theta_k$. There are $N - m$ entries along the $m$th diagonal, so the length-scale of discretization of the spherical harmonic is $1/(N - m)$.}\label{fig:sphere-harm}
\end{figure}

It is most instructive to analyze this system in the diagonal $Z$ basis, where $\hat{Y}_{jm}$ is only supported on the $m$th diagonal (see Fig. \ref{fig:sphere-harm}). It is clear from the structure of \eqref{eqn:msh-ham} that the $\hat{Y}_{jm}$ fermion mode will only be entangled with the $\hat{Y}_{j,-m}$ mode. Because the matrix entries of matrix spherical harmonics are a discretization of the usual commutative spherical harmonics $y_{jm}(\theta,\phi)$, we may evaluate traces of products of the $\hat{Y}_{jm}$ in terms of integrals of $y_{jm}(\theta,\phi)$ up to corrections suppressed by $O(1/(N - j))$.

\begin{figure}
\centering
\includegraphics[width=0.55\textwidth]{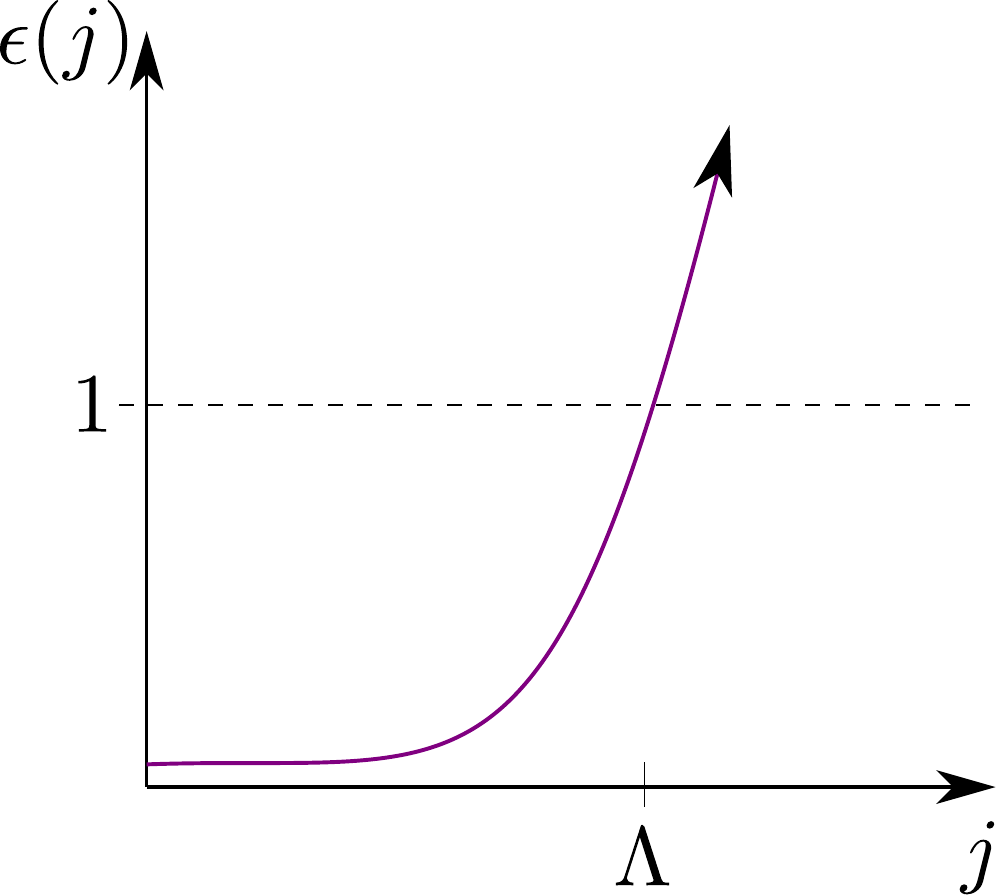}
\caption{The structure we take for the function $\epsilon(j)$ appearing in \eqref{eqn:msh-ham}. Specifying $\epsilon(j)$ is equivalent to specifying the couplings $g_i$ appearing in the derivative expansion of \eqref{eqn:full-ham}. We assume $\epsilon(j) \ll 1$ up to some cutoff scale $j=\Lambda$, after which it sharply transitions to $\epsilon(j) \gg 1$. From the point of view of the emergent noncommutative field theory, $\epsilon(j)$ is the spectrum of the kinetic term.}\label{fig:MTN-eps-graph}
\end{figure}

The entanglement structure of the ground state depends on the structure of $\epsilon(j)$. Modes $\hat{Y}_{jm}$ for which $\epsilon(j)$ is small will have large entanglement, and vice versa for modes for which $\epsilon(j)$ is large. We take $\epsilon(j)$ to have a form depicted in Fig. \ref{fig:MTN-eps-graph}: it is a growing function of $j$ that sharply transitions from much less than 1 to much greater than 1 at some cutoff $j=\Lambda$. As is standard on noncommutative geometries, a momentum scale $\Lambda$ is also a length-scale (see the discussion below \eqref{eqn:nonc-length-scale}). Physically, $\Lambda$ is the length-scale of extended objects which propagate along the fuzzy sphere geometry. As such, it is the scale of nonlocality of the theory, and we refer to it as the string scale. To ensure a local entanglement structure for our network we take $\Lambda \ll N$.

Unlike in \S\S\ref{sss:cc-qnc}, there are two distinct contributions to the entanglement structure of the ground state. $H_1$ couples (and therefore entangles) the $j,m$ and $j,-m$ matrix spherical harmonic modes, whereas $H_2$ couples entries of $\lambda$ modes \textit{along} the diagonals pictured in Fig. \ref{fig:sphere-harm}. To make contact with standard tensor networks, we wish for the former contribution to dominate. We do this by estimating the scaling of these two contributions in $N$ and $\Lambda$.

\begin{figure}[ht]
\centering
\includegraphics[width=0.5\textwidth]{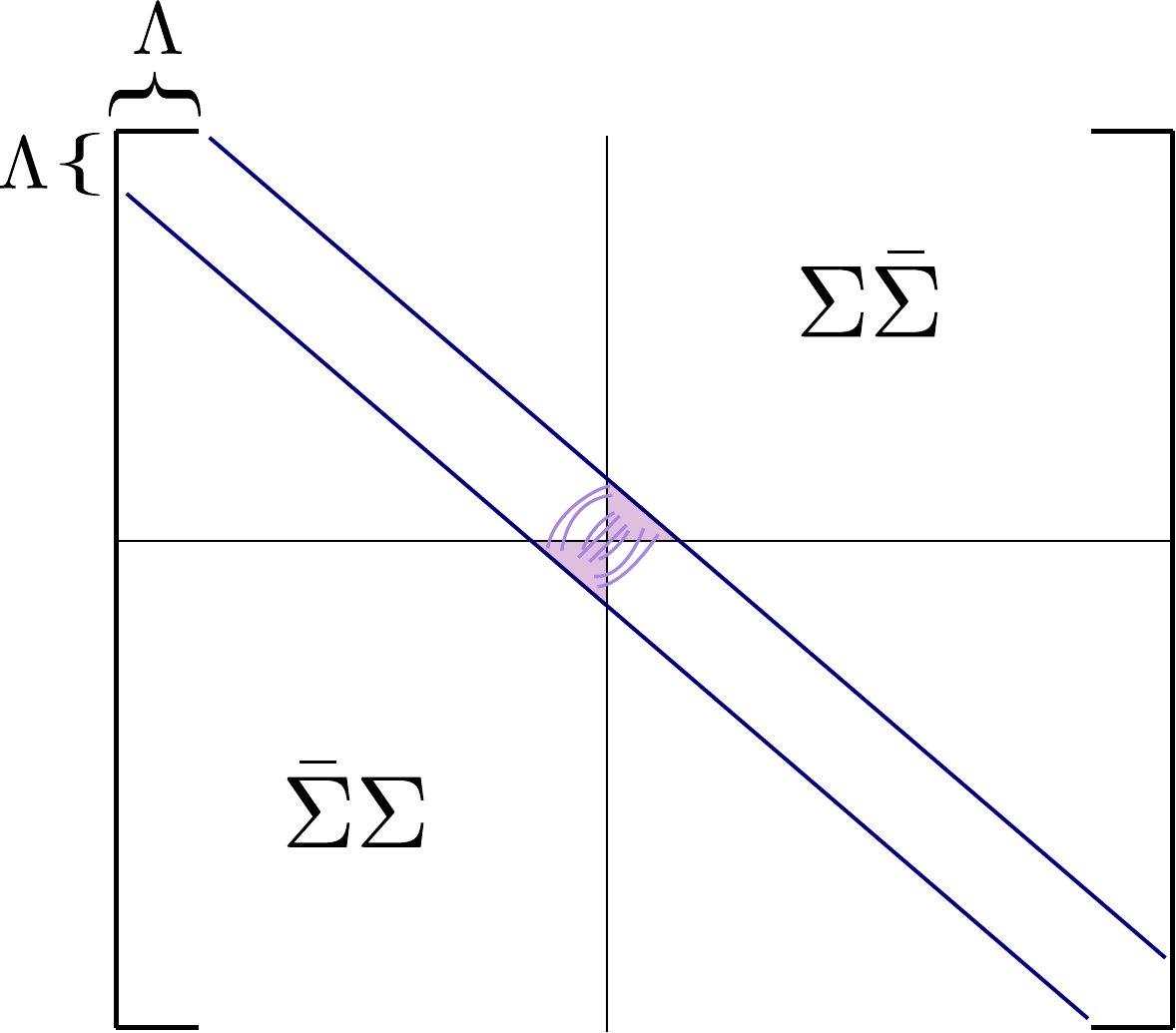}
\caption{In light of Figs. \ref{fig:MTN-partition} and \ref{fig:sphere-harm}, if we suppress all modes $\hat{Y}_{jm}$ with $j > \Lambda$ to their ground state then all excited qubits contributing to tensor network edges that cross the entanglement cut will be localized to the shaded corners of the $\Sigma \bar{\Sigma}$ and $\bar{\Sigma} \Sigma$ subblocks. In the parameter regime we are considering with $\Lambda \gg N^{1/2}$, this is the dominant contribution to the entanglement between the $\Sigma$ and $\bar{\Sigma}$ subregions of the network.} \label{fig:Lambda-graphic}
\end{figure}

The first issue is that the $\hat{Y}_{jm}$ modes are delocalized across the sphere. In view of Fig. \ref{fig:sphere-harm}, we look for a fourier-like transform that takes the delocalized form of spherical Harmonics along the matrix diagonal into a more local form. For a particular cutoff $\Lambda \ll N$, the $m$th diagonal will have $\Lambda - m$ activated modes entangled with the opposite diagonal. There are $N - m$ lattice sites along the $m$th diagonal, so we may squeeze each qudit to occupy $(N - m)/(\Lambda - m)$ lattice sites. The order of magnitude of the number of entangled qudits $N_q$ within the off-diagonal block (i.e. the number of cut tensor network edges contributing to the entanglement as in Fig. \ref{fig:Lambda-graphic}) therefore scales as
\begin{equation}\label{eqn:Nq-oom}
N_{q} \sim O(\Lambda^3 / N).
\end{equation}
On the other hand, the number of diagonals with excited degrees of freedom that cross the entanglement cut simply scales as $\Lambda$. This contribution to the entanglement is precisely what is computed in \cite{karczmarek2014entanglement}, and scales as $\Lambda \log \Lambda$. As we take $\Lambda \ll N$, note that we are within the area law regime found \cite{karczmarek2014entanglement}. We therefore wish to take the limit $\Lambda^3 / N \gg \Lambda$. In order to for the scale of nonlocality to be much smaller than the scale of the background geometry, we must take $\Lambda \ll N$. We therefore consider the parameter regime
\begin{equation}
N^{1/2} \ll \Lambda \ll N.
\end{equation}

\begin{figure}[ht]
\centering
\includegraphics[width=0.5\textwidth]{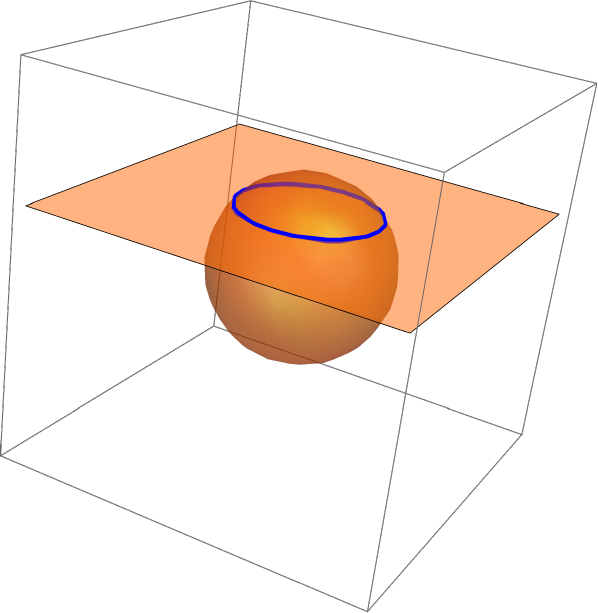}
\caption{A cap subregion of the fuzzy sphere, of the sort considered in \cite{karczmarek2014entanglement,frenkel2023emergent}. It is rotationally symmetric around the $z$ axis, and its boundary $\partial \Sigma$ (i.e. the entanglement cut) is simply a line of latitude at some constant $z$ value $z_c$. If the fuzzy sphere has radius 1, we may consider $z_c = 2M/N - 1$ for any integer $M$ such that $1 \leq M \leq N - 1$. To evaluate the entanglement, we simply need to count the number of tensor network lines as pictured in Fig. \ref{fig:network-apds-intro}(a) that cross $\partial \Sigma$.}\label{fig:cap-subregion}
\end{figure}

To evaluate the entanglement, we simply need to rotate to a basis where $\Theta_{\Sigma}$ induces a block decomposition of the matrix as in Fig. \ref{fig:MTN-partition} and count the number of qubits in the $\Sigma \bar{\Sigma}$ block that are entangled with their partners in the $\bar{\Sigma}\Sigma$ block. This is given by the sum
\begin{equation}\label{eqn:Nq-sum}
N_q = N_f \sum_{j=1}^{\Lambda}\sum_{m=-j}^j \left(\Tr[(\hat{Y}_{jm})^{\dag}_{\Sigma \bar{\Sigma}}(\hat{Y}_{jm})_{\bar{\Sigma} \Sigma}] + \Tr[(\hat{Y}_{jm})^{\dag}_{\bar{\Sigma} \Sigma}(\hat{Y}_{jm})_{\Sigma \bar{\Sigma}}]\right).
\end{equation}
As an instructive example we explicitly evaluate this sum in a simple the simple case where $\Sigma$ is a cap subregion on the sphere (see Fig. \ref{fig:cap-subregion}) so that $\Theta_{\Sigma}$ is diagonal in the same basis as $Z$. The spherical harmonics then take the form drawn in Fig. \ref{fig:sphere-harm}. The sum may be rewritten as
\begin{equation}
N_q = N_f \sum_{j=1}^{\Lambda}\sum_{m=1}^j \left(\Tr[(\hat{Y}_{jm})^{\dag} [\hat{Y}_{jm},\Theta_{\Sigma}]] - \Tr[(\hat{Y}_{j,-m})^{\dag} [\hat{Y}_{j,-m},\Theta_{\Sigma}]]\right).
\end{equation}
As the $\hat{Y}_{jm}$ form a complete basis for $N \times N$ matrices, we may expand the matrix $\Theta_{\Sigma}$ itself in terms of spherical harmonics:
\begin{equation}
\Theta_{\Sigma} = \sum_{jm}c_{\Sigma,jm}\hat{Y}_{jm}, \quad c_{\Sigma,jm} := \Tr[\hat{Y}_{jm}^{\dag}\Theta_{\Sigma}].
\end{equation}

Via the Moyal map \cite{Moyal:1949sk,Susskind:2001fb,Steinacker:2011ix} (and as strongly suggested by Fig. \ref{fig:sphere-harm}) we may approximate $c_{\Sigma,jm}$ up to $O(1/N)$ corrections as
\begin{equation}\label{eqn:cjm-areas}
c_{\Sigma,jm} = \int_0^{\pi} d\theta \int_s^{2\pi} d\phi \sin \theta d\theta d\phi\, \chi_{\Sigma}(\theta,\phi)y_{jm}(\theta,\phi) + O(1/N).
\end{equation}
The coefficients $c_{\Sigma,jm}$ are therefore simply determined by the properties of the Fourier decomposition of step functions. In particular, for large $j$ they have the structure
\begin{equation}
\sqrt{\sum_{m=-j}^j |c_{\Sigma,jm}|^2} \approx \frac{|\partial \Sigma|}{\pi j^2} + O(C/j^3).
\end{equation}
$C$ is the lengthscale of the radius of curvature of $\partial \Sigma$. Put simply, the projection of a characteristic function onto high-momentum (compared to the radius of curvature) fourier mode subspaces has magnitude proportional to the surface area (i.e. perimeter for the $2d$ case). For the case of the cap subregion  symmetry demands that only $c_{\Sigma,j0}$ is nonzero, so we simply have
\begin{equation}
|c_{\Sigma,j0}| \approx \frac{|\partial \Sigma|}{\pi j^2} + O(1/j^3).
\end{equation}
Plugging this back into \eqref{eqn:Nq-sum} we find
\begin{equation}\label{eqn:cap-area-law}
N_q \approx 2N_f |\partial \Sigma| \underbrace{\sum_{j=1}^{\Lambda}\sum_{m=1}^{j}\sum_{j'=1}^{N}\frac{1}{\pi {j'}^2}\Tr[\hat{Y}_{jm}^{\dag}[\hat{Y}_{jm},\hat{Y}_{j'0}]]}_{\text{A numerical coefficient independent of $\Sigma$}} +\, O(1/\Lambda).
\end{equation}
Taking $\Lambda$ large, we find that the number of tensor network legs partitioned by the entanglement cut is proportional to the perimeter length $|\partial \Sigma|$. As the entanglement is proportional to $N_q$, we find area law entanglement in this parameter regime.

We may extend this argument to more general subregions using the techniques of \cite{Frenkel:2023yuw}. Instead of $Z$ we diagonalize some other matrix $F$ (as in Fig. \ref{fig:network-apds-intro}), and take the radius of curvature $C$ to be given by $C \sim \sqrt{\Tr[F[X^i,[X^i,F]]]/\Tr[F^2]}$. It is important to take $C \ll \Lambda$. This matrix corresponds to the curvilinear coordinate $f(\theta,\phi)$ on the sphere, along with an orthogonal curvilinear coordinate $\phi_f$ which satisfies $[F,\cdot] \leftrightarrow -i\partial_{\phi_f}$. The computation was easy in the case of the cap subregion because spherical harmonics are eigenstates of $[Z,\cdot] \leftrightarrow -i\partial_{\phi}$. Because high-momentum spherical harmonics have the structure of plane waves on the surface of the sphere, we may find local rotations of the spherical harmonics that satisfy
\begin{equation}
\begin{split}
\tilde{y}_{jm}(f,\phi_f):= \sum_{m} c^f_{j,m,m'}(\theta,\phi)y_{jm'}(\theta,\phi), \quad \sum_{m''}c^f_{j,m,m''}(\theta,\phi) \bar{c}^f_{j,m',m''}(\theta,\phi) = \delta_{m,m'},
\end{split}
\end{equation}
such that $\tilde{y}_{jm}(f,\phi_f)$ is an eigenfunction of $\partial_{\phi_f}$. This definition may be promoted to the fuzzy sphere by promoting the $c^f_{j,m,m'}(\theta,\phi)$ to be functions of the $X^i$. The $\tilde{y}_{jm}(f,\phi_f)$ then get promoted to matrices $\tilde{Y}_{jm}$ that satisfy
\begin{equation}
\partial_{\phi_f} \tilde{y}_{jm}(f,\phi_f) = im\tilde{y}_{jm}(f,\phi_f) \leftrightarrow [F,\tilde{Y}_{jm}] = m\tilde{Y}_{jm}.
\end{equation}
This implies that $\tilde{Y}_{jm}$ now have the exact structure depicted in Fig. \ref{fig:sphere-harm}.

We now recall that the matrix commutator becomes the Poisson bracket $\{\cdot,\cdot\}$ on the fuzzy sphere surface. We therefore have
\begin{equation}
\begin{split}
&\{\tilde{y}_{j_1,m_1}(f,\phi_f), \tilde{y}_{j_2,m_2}(f,\phi_f)\} =\\
&=\sum_{m_1',m_2'}c_{j_1,m_1,m_1'}^f(\theta,\phi)c_{j_2,m_2,m_2'}^f(\theta,\phi)\{y_{j_1,m_1'}(\theta,\phi), y_{j_2m'_2}(\theta,\phi)\} + O(C/\min(j,j')),
\end{split}
\end{equation}
as the terms that dominate the Poisson bracket are those where the derivatives hit the $\tilde{y}_{jm}$ instead of the $c_{j,m,m'}$. Pulling this back to matix expression in turn implies
\begin{equation}
[\tilde{Y}_{jm},\tilde{Y}_{j'm'}] = \sum_{m_1',m_2'}c^f_{j_1,m_1,m_1'}(X^i)c^f_{j_2,m_2,m_2'}(X^i)[\hat{Y}_{jm},\hat{Y}_{j'm'}] + O(C/\min(j,j')).
\end{equation}
Using this expansion we may now run the exact same computation as before, finding again the result \eqref{eqn:cap-area-law} with the same $O(1)$ overall numerical coefficient:
\begin{equation}\label{eqn:gen-area-law}
N_q \approx 2N_f |\partial \Sigma| \sum_{j=1}^{\Lambda}\sum_{m=1}^{j}\sum_{j'=1}^{N}\frac{1}{\pi {j'}^2}\Tr[\hat{Y}_{jm}^{\dag}[\hat{Y}_{jm},\hat{Y}_{j'0}]] +\, O(C/\Lambda).
\end{equation}

\begin{figure}[ht]
     \centering
     \includegraphics[width=0.8\textwidth]{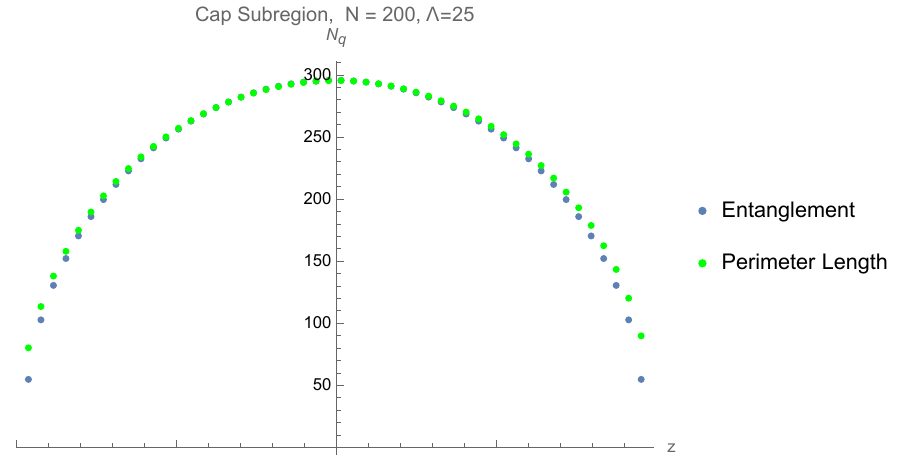}
     \caption{We numerically compute the sum that calculates $N_q$ in \eqref{eqn:Nq-sum} for the cap subregion (see Fig. \ref{fig:cap-subregion}) for various choices of $z$ at which we partition the sphere (blue dots). We compare the result to the functional form of $\sin(\arccos(z))$ -- the expression for the perimeter length $|\partial \Sigma|$ at each value of $z$ (green dots). Note that for small subregions the entanglement is smaller than what we would expect from the perimeter law. This occurs because the region area decreases much faster than the perimeter length, and when $\Sigma$ is of roughly the length-scale of nonlocality $\Lambda/N$ the entanglement saturates at a volume law.}\label{fig:cap-numerics}
\end{figure}

\begin{figure}[ht]
     \centering
     \begin{subfigure}[b]{0.45\textwidth}
         \centering
         \includegraphics[width=\textwidth]{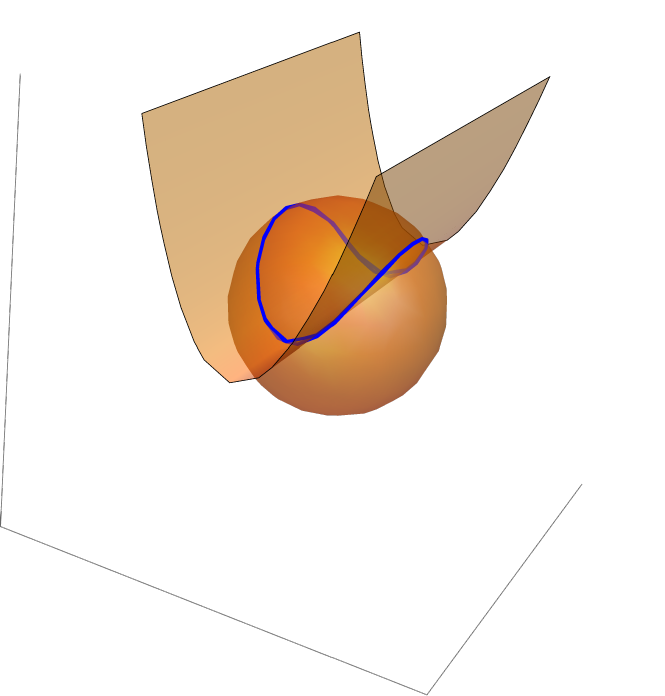}
         \caption{}
         \label{fig:2x2-sub-1}
     \end{subfigure}
     \hfill
     \begin{subfigure}[b]{0.45\textwidth}
         \centering
         \includegraphics[width=\textwidth]{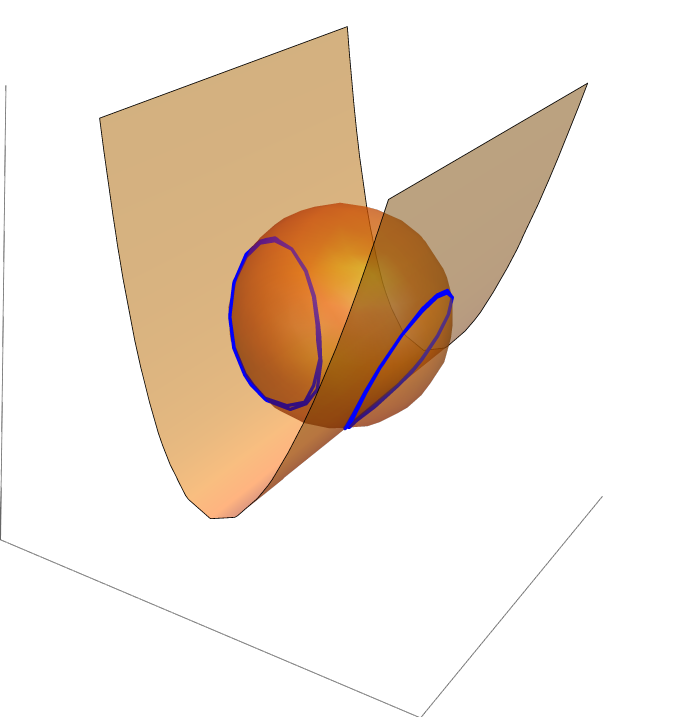}
         \caption{}
         \label{fig:2x2-sub-2}
     \end{subfigure}
     \caption{We consider a family of partitions parameterized by level sets of the function $f(\vec{x}) = z + x^2$. At $f(\vec{x}) = 1$, the subregion on the fuzzy sphere splits from one component (as in Fig. \ref{fig:2x2-sub-1}) to two disconnected components (as in Fig. \ref{fig:2x2-sub-2}). The matrix tensor network entanglement for this family of subregions is plotted against the perimeter in Fig. \ref{fig:2x2-numerics}.}\label{fig:2x2-cuts}
\end{figure}

\begin{figure}[ht]
     \centering
     \includegraphics[width=0.8\textwidth]{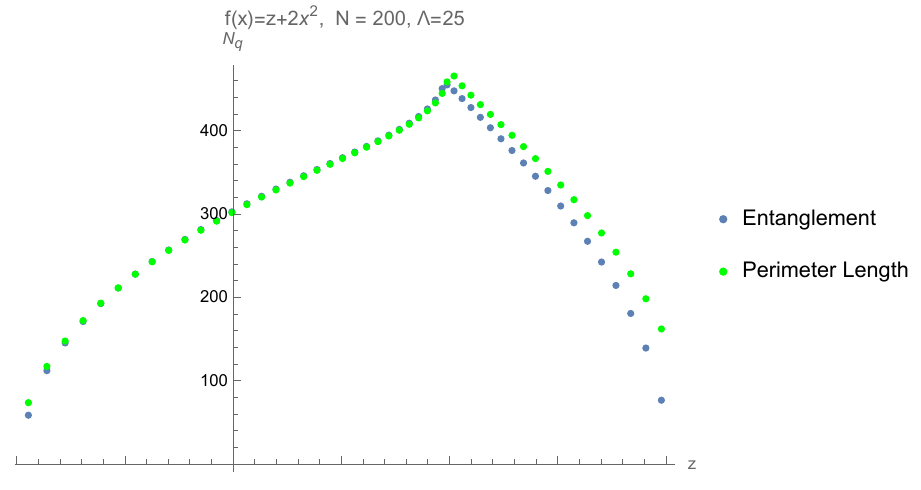}
     \caption{The numerical evaluation of \eqref{eqn:Nq-sum} for the family of entanglement cuts depicted in Fig. \ref{fig:2x2-cuts}. The cusp at $z=1$ is precisely the point at which the subregion separates into two (i.e. the transition from the regime in Fig. \ref{fig:2x2-sub-1} to the regime in Fig. \ref{fig:2x2-sub-2}. The overall coefficient in the match between entanglement and perimeter is identical to that in Fig. \ref{fig:cap-numerics}. Note that after the region separates into two (i.e. to the right of the cusp) the entanglement is subleading compared to the perimeter law. This is attributed to two effects. Immediately after the region splits into two, the two disconnected subregions are very close to each other and there will be many nonlocal tensor network legs stretching between them (note that at $\Lambda = 25$, the scale of nonlocality is $25\%$ of the sphere radius.) As the two regions shrink, we enter the regime where the area law saturates into a volume law, and the volume to area ratio increases with the number of disconnected components of the subregion.}\label{fig:2x2-numerics}
\end{figure}

To help the reader trust these quick noncommutative function manipulations, we present numerical results. See Fig. \ref{fig:cap-numerics} for a direct numerical evaluation of \eqref{eqn:Nq-sum} for the cap subregion that confirms the area law behavior. See Figs. \ref{fig:2x2-cuts} and \ref{fig:2x2-numerics} for numerics for another family of entanglement cuts that includes topologically nontrivial subregions.

\subsection{Emergent Non-Abelian Gauge Fields on the Links}\label{ssec:gauge-fields}

The observation that tensor networks may incorporate emergent bulk gauge fields living on their links is nothing new \cite{Dong:2023kyr,Akers:2024ixq,Akers:2024wab,Qi:2022lbd}. In this section, we simply demonstrate that this construction naturally arises from the structure of MQM.

The $U(N)$ symmetry of the network acts in the adjoint representation on $\lambda_{nn'}$. Consider acting with a diagonal $U \in U(1)^N \subset U(N)$. This will rotate the matrix elements as
\begin{equation}
\lambda^{\alpha}_{nn'} \rightarrow \lambda^{\alpha}_{nn'}e^{i(\theta_n - \theta_n')}.
\end{equation}
Note that $\lambda^{\alpha}_{nn'}$ transforms precisely with the opposite phase as with its partner $\lambda^{\alpha}_{n'n}$. There is therefore a natural $U(1)$ emergent bulk gauge field that lives on the tensor network links. This is the same $U(1)$ of the emergent Maxwell (or Chern-Simons) theory that lives on the surface of the fuzzy sphere \cite{Iso:2001mg}. To make contact with \cite{Dong:2023kyr,Akers:2024ixq}, we promote the emergent $U(1)$ gauge theory to a $U(p)$ gauge theory in the standard way \cite{Iso:2001mg,Susskind:2001fb,Dorey:2016mxm} -- by stacking $p$ fuzzy spheres on top of one another. Specifically, we take our coordinate matrices to be
\begin{equation}
\tilde{X}^i := X^i \otimes \mathbb{1}_p,
\end{equation}
where the $X^i$ are an irreducible representation of the $\mathfrak{su}(2)$ algebra as in \eqref{eqn:fuzz-sphere}. We take $X^i$ to be $N \times N$ and $\tilde{X}^i$ to be $pN \times pN$. The structure of $\tilde{X}^i$ encourages us to introduce a block decomposition for the $\lambda_{nn'}$. We introduce the symbol $\mathfrak{l}_{kk'}$ to denote the $p \times p$ block
\begin{equation}
\mathfrak{l}_{kk'} = \begin{bmatrix}
\lambda_{pk + 1,pk' + 1} & \ldots & \lambda_{pk+1,pk'+ p}\\
\vdots & \ddots & \vdots\\
\lambda_{pk+p,pk'+1} & \ldots & \lambda_{pk+p,pk'+p}
\end{bmatrix}.
\end{equation}
We have introduced an index $k,k'$ which runs from $1$ to $N$, whereas $n,n'$ runs from $1$ to $pN$.

We plug $\tilde{X}^i$ into the Hamiltonian \eqref{eqn:full-ham}. Note that $\tilde{X}^i$ satisfy
\begin{equation}
[\tilde{X^i},U] = 0 \quad \text{for} \quad U \in U(p)^N.
\end{equation}
This Weyl-like subgroup of $U(pN)$ acts on the $p \times p$ $\mathfrak{l}_{kk'}$ blocks as
\begin{equation}
\mathfrak{l}_{kk'} \rightarrow U_k \mathfrak{l}_{kk'}U_{k'}^{\dag}, \quad U_k,U_{k'} \in U(p).
\end{equation}
We have therefore promoted our tensor network to one where the links are $U(p)$ bifundamentals as opposed to $U(1)$ bifundamentals, and have found an emergent $U(p)$ gauge theory. The structure of \eqref{eqn:full-ham} ensures that $\mathfrak{l}_{kk'}$ only appear in the Hamiltonian in the combination $\Tr[\mathfrak{l}_{kk'}^{\dag} \mathfrak{l}_{k'k}]$, $\Tr[\mathfrak{l}_{kk'} \mathfrak{l}_{k'k}]$, or $\Tr[\mathfrak{l}_{kk'}^{\dag} \mathfrak{l}_{k'k}^{\dag}]$. The ground state is therefore invariant under the larger group $U(p)^{N^2}$ that acts as
\begin{equation}
\mathfrak{l}_{kk'}, \mathfrak{l}_{k'k} \rightarrow U_{kk'} \mathfrak{l}_{kk'}U_{kk'}^{\dag}, U_{kk'} \mathfrak{l}_{k'k}U_{kk'}^{\dag}, \quad U_{kk'} \in U(p).
\end{equation}
The Hilbert space of $\mathfrak{l}_{kk'}$ decomposes into $U(p)$ irreps. By standard arguments \cite{ghosh2015entanglement}, in the ground state the irreps of $\mathfrak{l}_{kk'}$ are maximally entangled with the corresponding irreps of $\mathfrak{l}_{k'k}$.

\section{Discussion and Outlook}\label{sec:disc}

We have introduced a new family of tensor network states inspired by MQM and noncommutative geometry. They appear to naturally include notions of background independence, area-preserving diffeomorphism invariance, emergent gauge theory, and a string scale. We have demonstrated that despite the nonlocality UV/IR mixing endemic to noncommutative field theories, the low-energy tensor network states exhibit area law entanglement.

\begin{figure}[ht]
\centering
\includegraphics[width=0.6\textwidth]{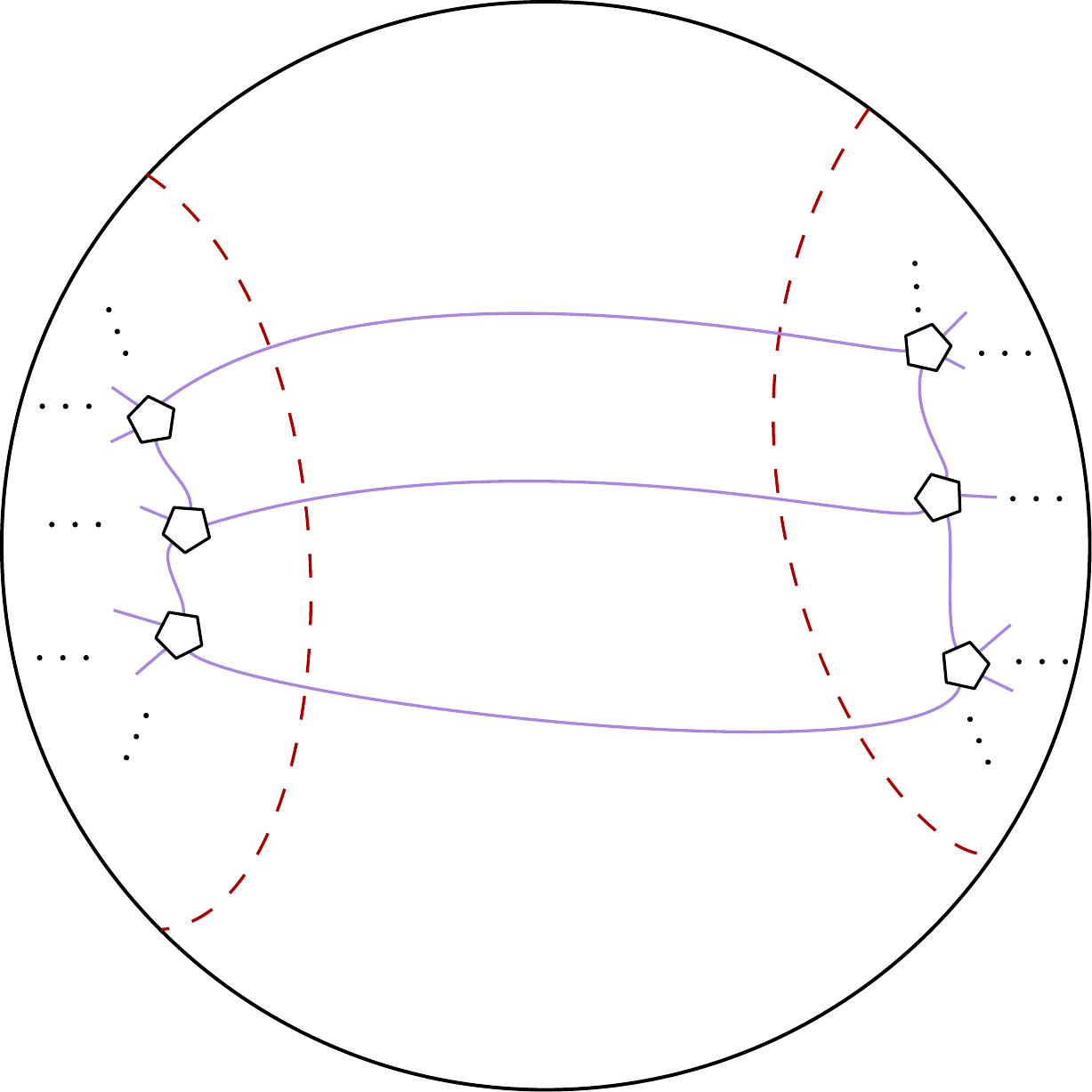}
\caption{Consider a tensor network on a fuzzy AdS spacetime -- perhaps a noncommutative analog of the HaPPY code \cite{pastawski2015holographic}. The tensor network states we consider naturally have very nonlocal links contributing to the entanglement structure. This nonlocal entanglement serves to pull RT surfaces closer to one another. It would be interesting to check if the ground state of Hamiltonians like \eqref{eqn:full-ham} may induce the appropriate backreaction between RT surfaces to match the behavior in the bulk of AdS/CFT.}\label{fig:nonlocal-network}
\end{figure}

It would be interested to explicitly check the RT surface and bulk reconstruction properties of these networks, especially on fuzzy AdS spaces, perhaps using the methods of \cite{Cheng:2022ori,ffhs:2024xx}. In particular, it would be interesting to make sense of tensor network links that might stretch far across the geometry. These modes suggest a natural way to encode mutual information between distant subregions, perhaps inducing backreaction between RT surfaces (see Fig. \ref{fig:nonlocal-network}). One could also ask whether the dynamical background independence we consider in this work is identical to the notion of background independence explored in \cite{Akers:2024ixq}.

It would also be interesting to explore whether such matrix tensor networks are the appropriate language in which to capture $\alpha'$ (or general higher-derivative) corrections to the entanglement entropy (as in \cite{Dong:2013qoa,Wall:2015raa}). More generally, it is an interesting question whether a matrix tensor network on a fuzzy $S^5$ background (perhaps using the results of \cite{Fiore:2022twy}) is the correct way to capture how the compact directions of AdS/CFT may be incorporated into tensor networks. Similarly, it is interesting to check whether these networks accurately capture the ground-state entanglement behavior of BFSS.

Lastly, although we have proposed a manner if incorporating spatial APDs into tensor networks, it is interesting whether time reparameterizations may also be incorporated into such a context, e.g. by using constructions such as \cite{Banados:2001xw}. Alternatively, it is possible that the dynamics generated by \eqref{eqn:full-ham} or similar Hamiltonians might be related in some way to bulk time evolution. \cite{ffhs:2024xx}

\section*{Acknowledgements}

I am partially supported by the NSF GRFP under grant no. DGE-165-651. I am indebted especially to Xiao-Liang Qi and Aditya Cowsik for extensive insightful discussions. I am also grateful to Ronak M. Soni, Watse Sybesma, Laurens Lootens, Chris Akers, Jason Pollack, and Annie Y. Wei for extensive comments on an early draft. I would like to thank Annie Y. Wei in particular for discussion that inspired the addition of \S\S\ref{ssec:gauge-fields}. I would also like to thank DAMTP and the University of Cambridge for hospitality during a significant portion of this work.

\appendix

\bibliographystyle{JHEP}
\bibliography{refs}
 
\end{document}